\documentclass{pinchcr}   
\newcommand{\stylename}{\texttt{pinchcr}}

\usepackage{graphicx}
\usepackage{dcolumn}
\usepackage{bm}
\usepackage{amssymb}
\usepackage{multirow}
\usepackage{color}
\usepackage{hyperref}
\usepackage{ulem}
\usepackage{float}

\title{Phases of active matter in two dimensions
}

\author{Leticia F. Cugliandolo}
\affiliation{Sorbonne Universit\'e, \\
Laboratoire de Physique Th\'eorique et Hautes Energies, CNRS UMR 7589\\
Tour 13, 5\`eme \'etage, 
4 Place Jussieu, \\
75252 Paris Cedex 05,  France}

\author{Giuseppe Gonnella}
\affiliation{Universit\`a  degli  Studi  di  Bari  and  INFN Sezione  di  Bari,  \\
Dipartimento di Fisica, \\
via  Amendola  173, \\
I-70126  Bari,  Italia}
\authors{2}

\authors{2}

\begin{document}

\maketitle

\preface
These notes focus on the description of the phases of matter in two dimensions.
Firstly, we  present a brief discussion of the phase diagrams of bidimensional 
interacting passive systems,
and their numerical and experimental measurements. 
The presentation will be short and  schematic.  We will 
complement these notes with a rather complete bibliography 
that should guide the students in their study of the development of 
this very rich subject over the last century.
Secondly, we summarise very recent results on the 
phase diagrams of active Brownian disks and active dumbbell systems in two dimensions. 
The idea is to identify all the phases and to relate, when this is possible, 
the ones found in the passive limit with the ones observed at 
large values of the activity, at high and low densities, and for both types of constituents. 
Proposals for the mechanisms leading to these phases will be discussed. The  
physics of bidimensional active systems 
open many  questions, some of which will be listed by the end of the Chapter.

\acknowledgements

We want to warmly thank our collaborators on studies of active matter; in alphabetical order
Mathias Casiulis, Olivier Dauchot, Pasquale Digregorio, Gianluca Laghezza, 
Antonio Lamura, Demian Levis, Davide Loi, Davide Marenduzzo, Alessandro Mossa,
Stefano Mossa, Ignacio Pagonabarraga, 
Isabella Petrelli, Antonio Suma, and Marco Tarzia, 
from whom we learnt most of what we present in these 
notes. Leticia F. Cugliandolo is a member of Institut Universitaire de France. She thanks the KITP University of Santa Barbara at 
California (National Science Foundation under Grant No. NSF PHY17-48958) where these lectures notes were prepared in part. Our numerical work
was possible thanks to allocated time at 
the MareNostrum Supercomputer at the Barcelona Supercomputing Center (BSC), the 
IBM Nextscale GALILEO at CINECA (Project INF16- fieldturb) under CINECA-INFN agreement and Bari Re-CaS e-Infrastructure 
funded by MIUR through PON Research and Competitiveness 2007-2013 Call 254 Action I.

\tableofcontents

\maintext

\chapter{Phases of planar active matter}

\section{Introduction}

\label{sec:introduction}

These notes describe the content of a two-lecture course given by L. F. Cugliandolo at the summer school 
``Active Matter and  Non Equilibrium Statistical Physics'' held at the Les Houches School of Physics 
(August-September 2018). They will be published in a volume of the Les Houches collection dedicated to this school. 
A detailed introduction to active matter will appear  in other Chapters of this book and we will not cover it here. 

Our plan is to explain some of the peculiarities of passive and active matter in two dimensions.
While under most natural conditions matter fills three dimensional volumes, systems 
can be restrained to occupy one or two dimensional spaces with convenient confining potentials. Low 
dimensional systems are interesting for practical and conceptual reasons.
First of all, they are realised in Nature, and
some classical and quantum examples are colloidal suspensions under confined conditions, liquid crystal, 
magnetic and superconducting films, and electrons trapped at liquid helium surfaces, to name just a few.
Secondly, they can be easier to study numerically and experimentally than their higher dimensional extensions. 
But most importantly, they can pose specific and interesting questions of fundamental  relevance. We will 
discuss some of these in these notes.

In the rest of the Introduction we briefly describe some salient features of equilibrium matter in $2d$,
and the problems that the injection of energy poses on the analytic treatment 
of  far from equilibrium systems. We will also state which are the concrete issues that we addressed in the 
lectures and that we will treat in the body of these notes.

\subsection{Equilibrium: role of symmetries and space dimension}

Let us start with an {\it aper\c{c}u} of passive matter in low dimensions.
In equilibrium, the interactions are responsible for the richness and complexity of the phases in which matter can exist. 
Over the last century a rather good understanding of some of these phases, namely the solid, liquid and gas, 
and a partial understanding of more exotic cases, such as glasses or plasmas, have also been reached. However, the 
phase diagram of matter in two dimensions is still under debate. While it is well-established that crystals cannot exist in 
two dimensions, solids (with a non-vanishing shear modulus\footnote{The shear modulus quantifies the deformation caused 
under a force $F$ parallel to one of the material's surfaces while its opposite face experiences an opposing force such as friction. It is defined as 
the ratio of shear stress over shear strain, $(F/A)/(\Delta x/l)$, with $A$ the area of the surface, $\Delta x$ the displacement caused 
by the force and $l$ the original length. Fluids have vanishing shear modulus.}) do. 
However, the mechanisms driving solid's melting 
and the transition towards the liquid phase are not fully settled yet.

Symmetries are one of the important factors that determine the collective behaviour of an equilibrium system.
Two basic symmetries of the laws of nature are {\it translational} and {\it rotational} invariance. These are consequences 
of the fact that, typically, in systems with pairwise interactions the potential energy depends on $r_{ij} = |{\bf r}_i - {\bf r}_j|$, 
with  ${\bf r}_i$ and ${\bf r}_j$ the positions of the two constituents, 
and not on ${\bf r}_i$ and ${\bf r}_j$ independently.
While the state of a macroscopic system in equilibrium is expected to respect both at sufficiently high temperature and low density,  
in the opposite cold and/or dense limit one or both of these symmetries can be broken, and macroscopic observations may, 
for instance, depend on the orientation of the sample. A phase transition then reflects the failure of the system to 
respect the symmetry of its Hamiltonian.

In a crystalline phase, the constituents order in a perfect periodic and stable array that covers the sample. Translational invariance is 
broken since the lattice is 
periodic under  translations by discrete vectors only. Moreover, the orientation of the local crystallographic axes,  
associated to each particle {\it via} its first neighbours, is always the same, and rotational invariance is also broken.
In a gas or a liquid both translational and rotational symmetries are respected, the constituents are randomly placed in the 
samples, and these are completely isotropic. In between these two extremes, there are situations in which the 
systems can be partially ordered, for example, exhibiting long-range orientational order but only short-range positional one, 
a case that we will explore here.

In low dimensional systems with short-range interactions translational order is forbidden by the so-called Mermin-Wagner theorem
but a solid phase with quasi long-range translational order is allowed.
A proposal for the mechanism for the transition from solid to liquid led by the 
dissociation of dislocation pairs\footnote{A dislocation is a line, plane, or region in which there is a discontinuity in the normal lattice structure of a crystal.}
was proposed by Kosterlitz \& Thouless (KT) in their 1972 \& 1973 seminal papers. 
However, knowing that long-range orientational order  is possible in $2d$~\cite{Mermin68},~\shortciteN{Nelson1979} 
and  \shortciteN{Young1979} 
modified the KT picture and claimed that the transition actually occurs in two steps, the second one being linked to the 
unbinding of disclinations.\footnote{A disclination is a line defect in which rotational symmetry is violated.} 
In this picture the intermediate phase keeps quasi long-range orientational order, 
allowing the system to increase its entropy over the solid at a moderate energy cost. The two transitions are 
of infinite order, {\it \`a la} KT, and the full description is commonly called the KTHNY scenario. Part of the latter picture 
has been recently contested in passive systems with 
hard potentials (Bernard \& Krauth, 2011) 
and we will review the modifications proposed in the body of the notes.

\subsection{Out of equilibrium: lack of generic guiding principles}

In out of equilibrium active systems, the injection of energy adds complexity and richness 
to the variety of behaviours that a large system can have.
In particular, one could revisit the no-go theorems for crystalline order in two dimensions, one could wonder  whether 
the solid phases still exist and whether they are favoured or disfavoured under activity, or whether disorder phases are enhanced by the injection of 
energy. All these are questions that merit attention. 

Much of the studies of active matter systems have focused on low density limits in which solid phases do not exist. 
Powerful analytical approaches, such as hydrodynamic theories, and numerical ones, such as Lattice Boltzmann methods, 
can be developed and successfully applied to describe the behaviour in dilute limits. 
These are treated in detail in other Chapters of this book. In contrast, less has been 
done for dense systems, cases in which  the passive limit is harder to unveil.

Another difficulty, or richness, of out of equilibrium systems is that
{\it thermodynamic concepts} have to be revisited as they are not necessarily defined. 
{\it Effective temperatures} and chemical potentials, intensive parameters in a thermodynamic approach, 
have been successfully used in the context of glassy physics, see {\it e.g.} the review article~\cite{cugl:review},
but become dynamic concepts that need to be measured carefully, 
separating time-scales and taking into 
account possible strong spatial heterogeneities.
The definition of pressure, that appears linked to density and temperature in the equilibrium equations of state, and 
allows one to estimate phase diagrams, also needs to be revisited out of equilibrium. 
Indeed, as it has been noticed in several occasions, the mechanical and thermodynamic 
definitions that are equivalent in equilibrium are not necessarily so out of equilibrium
and the mere existence of an equation of state becomes an issue in itself.

Confronted with the difficulty of deriving analytic results for interacting many-body systems,
{\it numerical methods} can come to our rescue and help us  understanding at least some aspects of 
the collective behaviour of matter, especially under dense conditions and the effects of activity. 
In the numerical studies of the equilibrium properties of 
passive systems one has the freedom to choose between {\it molecular 
dynamics simulations}, in which Newton equations of motion are integrated over sufficiently long time scales or, {\it Monte Carlo simulations} in which non-physical transition probabilities are sometimes 
chosen to optimise the sampling of  the equilibrium measure. In the context of active matter, the numerical integration of the  
actual dynamical equations is preferred [see, however, Levis and Berthier~\citeyear{Levis-Berthier} and Klamser 
{\it et al.}~\citeyear{KlKaKr18} for kinetic Monte Carlo methods for active systems]
putting a computational limit to the size and time scales that can be studied.

\subsection{These lectures}

Concretely, the aim of these two lectures has been: (1) to expose the students to a classical problem in statistical physics, the 
one of order and disorder in two dimensional systems (possible glassy aspects have been intentionally left aside); 
(2) to introduce them to  some standard tools used in the study of molecular systems;
(3) to discuss some aspects of dense, dry, interacting active matter confined to a plane. The latter system combines the 
difficulties of low-dimensional passive matter and the effects introduced by the constant injection of energy, partially dissipated
and partially used by the system itself. 
It is a challenging kind of problem with many open routes for further study, some of which are discussed by the 
end of the notes.

\section{Models and observables}

Numerous models of active matter exist
in the literature but we will not explain them all here. We will instead focus on some  that are directly inspired by 
the microscopic modelling of atomic and molecular passive systems, and hence admit a clear and simple equilibrium limit.
Such agent based active matter models make choices on (1) the form of the constituents, (2) the inter-particle interactions, 
(3) the coupling to the solvent or medium in 
which the agents are immersed, (4) the way in which the activity acts, and (5) the dimension and form of the confining space. 
In Sec.~\ref{subsec:models} we  list and discuss some common  choices made on these five properties. 
Besides, in Sec.~\ref{subsec:observables}
we define and discuss some observables that serve to characterise order 
(or the lack of it) in molecular systems, and that we will use to find the phases of the 
active problem.

\subsection{Models}
\label{subsec:models}

\subsubsection{The form of the constituents}

Playing with individual spheres one can switch from ``atomic'' to ``molecular'' models. More precisely, the constituents can be 
chosen to be simply spherically symmetric objects ({\it disks} in two dimensions), they can be linked together with a spring or a 
rod to form a {\it dumbbell} or, one can join many of them to build a {\it polymer}. As well-known in passive liquid equilibrium 
and out of equilibrium studies, the behaviour of an ensemble of such various objects can be rather different. Important shape 
effects have also been found in active matter and we will discuss some of them here.

Specifically, a disk is a spherically symmetric planar object with mass $m_{\rm d}$ and diameter $\sigma_{\rm d}$. A dumbbell is
a diatomic molecule made of two such disks linked together with a massless spring or a rigid rod. There is  
a head-tail nature attached to each disk. Adding more such monomers one can also build polymers of any desired length
with adequate bending properties.

\subsubsection{Inter-particle interactions}

\begin{figure}
\vspace{-2cm}
\begin{center}
\includegraphics[scale=0.3]{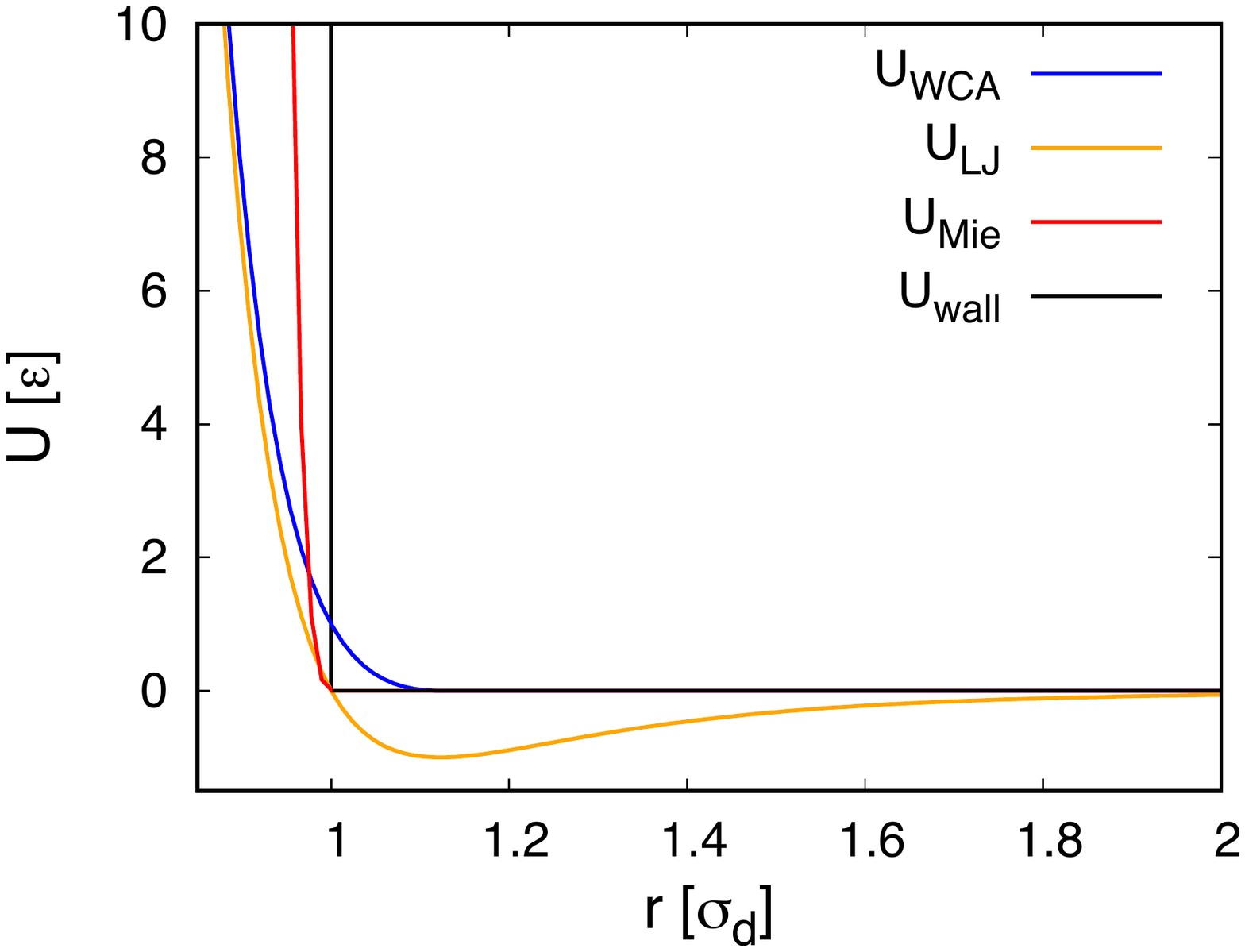}
\end{center}
\vspace{-2.25cm}
\caption{A sketch of the Lennard-Jones, WCA, a hard wall and a truncated and shifted Mie potential, see the text for their definitions.
}
\label{fig:potentials}
\end{figure}

In atomic systems the potential typically has an attractive tail due to Van der Waals dipole-dipole
interactions that decays as a power law 
$-1/r^{6}$ at large separations $r$ while at short distances the overlap of the electron clouds 
makes the potential strongly repulsive. The minimum, located  at a distance of a few Angstroms with a value of a hundred Kelvin, 
is responsible for the formation of crystals. This kind of potential is also used to 
describe the interactions between colloidal particles in a solvent although the length and energy scales involved can be very different from the 
atomic ones (say, length scales between a nanometer and a micrometer).
In some cases, only the repulsive part that accounts for the excluded volume interaction between colloids is retained. 
Without entering into a detailed justification for the latter choice, 
it is commonly adopted in the context of active 
matter as well. The potential is then taken to be a generalised Lennard-Jones (LJ) of Mie type~\cite{Mie1903}, truncated and shifted \`a la Week-Chandler-Andersen (WCA, 
\citeyearNP{Weeks}).
Calling ${\bold{r}}_i$ the $d$-dimensional position of the center of the $i$th particle and 
$r=|{\bold{r}}_i-{\bold{r}}_j|$ the inter-particle distance, 
the short-ranged repulsive potential takes the form
\begin{eqnarray}
U(r) =
\left\{
\begin{array}{ll}
4\varepsilon \, [({\sigma}/{r})^{2n}-({\sigma}/{r})^{n}]+\varepsilon \qquad & \qquad  \mbox{if} \quad  r< \sigma_{\rm d}=2^{1/n}\sigma
\; , 
\vspace{0.1cm}
\\
0 & \qquad  \mbox{otherwise}
\; ,
\end{array}
\right.
\end{eqnarray}
with $n$ a parameter that tunes the softness of the potential ($n=6$ is the usual LJ one and increasing $n$ the potential gets harder), 
$\epsilon$ an energy scale and $\sigma$ a length scale chosen to be of the order of the particle diameter. The hard wall, LJ, Mie and WCA 
potentials are compared in Fig.~\ref{fig:potentials}.

\subsubsection{Coupling to the environment}

The dissipation and fluctuations induced by the coupling to the environment are commonly mimicked {\it \`a la} Langevin, 
by adding friction and noise terms to the equations of motion. After inspection of the different time-scales involved in the 
motion of the active (Brownian) particles and the constituents of the environment, a Markovian (no memory) assumption is
usually justified. The noises are then chosen to be Gaussian, with zero mean and delta correlations proportional to 
the friction coefficient $\gamma_{\rm d}$ and the temperature of the bath $T$:
$\langle \xi^a_{i}(t) \, \xi^b_{j}(t') \rangle = 2 \gamma_{\rm d} k_B T \delta_{ij} \delta(t-t') \delta_{ab}$, with $a,b=1,\dots,d$ and $d$ the dimension of space.
The Boltzmann constant is denoted $k_B$.

\subsubsection{Activity}

The active forces can be  persistent or not. Persistency means that the 
constituents have a preferred direction along which the active force is permanently applied.
For instance, in the dumbbell models, the molecules are elongated along a 
main axis, with each bead having a head or tail nature. The active force is always exerted along this 
axis from tail to head~\shortcite{Cugliandolo2017,Petrelli2018}. An example with non-persistent activity 
is the one treated in Loi {\it et al.}~\citeyear{cugl-mossa2}, a polymeric model in which the 
active force was applied on temporal intervals of a given duration only, on a preselected monomer, and with 
random direction. In the case of the active Brownian disks, these are assumed to have a direction 
attached to them, along which the active force is always applied~\cite{Henkes11,Fily12,Redner13,Fily14,Redner2016classical}.

\subsubsection{The  P\'eclet number}

The P\'eclet number is an 
adimensional parameter 
defined as Pe = $F_{\rm act} {\sigma_{\rm d}}/(k_BT)$ 
that quantifies the strength of the activity as compared to the 
thermal fluctuations. It has a twofold interpretation. One is  as a ratio between the advective transport, say
$\ell v = F_{\rm act} \sigma_{\rm d} /\gamma_{\rm d}$ and the diffusive transport $D = k_BT/\gamma_{\rm d}$. The 
other one is as the ratio between the work done by the active force when translating the disk by its 
typical dimension, $F_{\rm act} \sigma_{\rm d} $, and the thermal energy, $k_BT$.
The P\'eclet number can be tuned by changing $F_{\rm act}$ at fixed $\gamma_{\rm d}$ and $k_BT$, 
for example.

\subsubsection{Reynolds number}

The Reynolds number confronts the strength of the inertial and viscous  forces
Re = $\rho L v/\mu = (\sigma_{\rm d} F_{\rm act}/\gamma_{\rm d})/(\gamma_{\rm d} \sigma_{\rm d}^2/m) = 
m_{\rm d} F_{\rm act}/(\gamma^2_{\rm d} \sigma_{\rm d})$ and it is typically very small, say $10^{-2}$, for the 
values of the parameters used in the simulations.

\subsubsection{The active Brownian disks equations of motion}

The dynamics of active Brownian disks is usually studied in the over-damped limit. 
The particles self-propel 
under a constant  modulus force 
$F_{\rm act}$ along a (rotating) direction $\bold{n}_i=( \cos{\theta_i(t)},\sin{\theta_i(t)})$  attached to the particle with $\theta_i$ the 
angle formed with a fixed axis. The disk center positions obey
\begin{equation}
	\gamma_{\rm d}\dot{\bold{r}}_i=F_{\rm act} \bold{n}_i- {\boldsymbol{\nabla}}_i\sum_{j(\neq i)}U(r_{ij}) + \bm{\xi}_i \; ,\qquad\qquad \dot{\theta}_i=\eta_i \; . 
	\label{eq:active-BD}
\end{equation}
The angular noise is taken to be Gaussian with zero mean and correlations $\langle \eta_i(t) \eta_j(t') \rangle = 2D_\theta \delta_{ij}
\delta(t-t')$.
The units of length, time and energy are given by  $ \sigma_d$, $\tau= D^{-1}_\theta$ and $\varepsilon$, respectively. 
The angular diffusion coefficient is commonly fixed to $D_\theta = 3k_BT/(\gamma_{\rm d} \sigma^2_d)$ (apart from the 
unimportant numerical factor that depends on the form of the objects, this is also the angular diffusion coefficient of a single active 
dumbbell).

\subsubsection{The active dumbbells equations of motion}

We model the time evolution of each sphere in the dumbbell through a Langevin equation of motion that acts on the 
position of the centre of each bead, ${\mathbf r}_i$, and is given by
\begin{equation}
m_{\rm d} \ddot {\mathbf r}_i=-\gamma_{\rm d}\dot {\mathbf r}_i -{\boldsymbol \nabla}_i \sum_{j(\neq i)} U(r_{ij})+
{\mathbf F}_{\rm act}+
 {\boldsymbol \xi}_i(t) \; ,
\label{bd}
\end{equation}
where $i=1,\dots, 2N$ is the sphere index and 
${\boldsymbol \nabla}_i=\partial_{\mathbf{r}_i}$.  Since the noises acting on the two beads that form a dumbbell 
are independent, the combined stochastic force can make the dumbbells rotate.

As usual in the Langevin description of the random motion of particles and mol\-e\-cules, in cases in which the inertial
time-scale $m_{\rm d}/\gamma_{\rm d}$ is much 
shorter than all other interesting time-scales in the problem, one can take an over-damped limit and basically 
drop the term $m_{\rm d} \ddot {\bf r}_i$ from the equations of motion. In this limit, the equations of motion are first order in the time derivative, as in Eqn.~(\ref{eq:active-BD}), but the form of the molecules and in particular the (averaged) length between the colloidal 
centres will have a notable importance in the collective behaviour of the macroscopic system.

\subsubsection{Space and packing fraction}

The constituents are typically confined to displace in a $d$ dimensional box with linear length $L$ and 
periodic boundary conditions, a situation that minimizes finite size effects and 
evades the difficulty of modelling the interactions with the boundary walls. The packing fraction is then 
defined as $\phi =\pi{\sigma^2_d}N/(4L^d)$ (with an extra factor of 2 for the dumbbells, 
for which the distance between the centres of the two coloids 
is fixed to $\sigma_{\rm d}$) 
and can be tuned by changing $N$ or $L$, for example.

\subsection{Observables}
\label{subsec:observables}

In this Section we define a number of observables and correlation 
functions that are used to 
characterise positional and orientational order in molecular systems.
In all the definitions we keep the time-dependence to evaluate the 
possible evolution of these quantities and thus quantify the 
dynamics of the problem.

\subsubsection{Voronoi tessellation}

\begin{figure}
\hspace{1.5cm} (a) \hspace{2.8cm} (b) \hspace{2.8cm} (c)
\begin{center}
\includegraphics[scale=0.21]{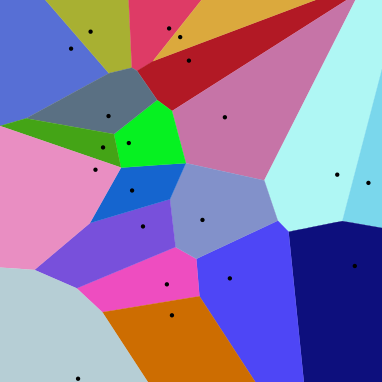}
\hspace{0.25cm}
\includegraphics[scale=0.5]{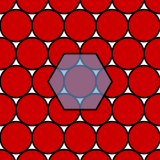}
\includegraphics[scale=0.31]{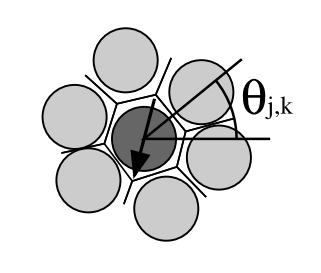}
\end{center}
\caption{(a) An example of Voronoi tessellation. (b) Close packing of disks, the hexagonal lattice.
(c) The definition of the angle $\theta_{jk}$ between a bond and a reference (horizontal) axis. The local 
hexatic order parameter represented as a vector attached to the central particle. The two last panels have been borrowed from E. Bernard's PhD thesis.}
\label{fig:voronoi-hexagonal}
\end{figure}

The first issue we want to resolve is to attribute a notion of neighbourhood to the configurations.
This is done with the help of a Voronoi tessellation. 
In our problem, we start by assigning $N$ points on the plane to the centres of the disks, be them the 
individual particles or the beads building the diatomic molecules. We then 
partition space in areas such that each space point is closer to the centre of one disk 
than to any other one. The borders of the areas thus constructed are polygons and the number of neighbours of a point 
is given by the number of sides of the polygon, see the sketch in Fig.~\ref{fig:voronoi-hexagonal}~(a). 
An example of a concrete 
dumbbell configuration with its Voronoi tessellation is shown in Fig.~\ref{fig:tessellation}~(a). 

\subsubsection{Fluctuating local density}

One can measure {\it local densities} $\phi_j$, with a discrete index dependence, 
in at least two ways, and construct with them  histograms and 
probability distribution functions.

With the first method, for each bead, one first estimates the local density as the ratio between its surface  
and the area of its Voronoi region $A_j^{\rm Vor}$. This value can then be coarse-grained by averaging the single-bead densities over 
a disk with radius $R$ around the concerned bead, 
\begin{equation}
\phi_j(t) = \frac{1}{n_R^{(j)}(t)} \sum_{i \in S_R^{(j)}} 
\frac{\pi \sigma_{\rm d}^2}{4 A_i^{\rm Vor}(t)}
\; . 
\end{equation}
where $n^{(j)}_R$ and $S^{(j)}_R$ are the number of particles and the
spherical disk with radius R around the $j$th bead, respectively.

With the second method,
one constructs a square grid on the simulation box, for each point in the grid one calculates a coarse grained 
local density $\phi_j$ over a circle of given radius $R$, and one finally assigns this density value  to the grid point.

\begin{figure}[h!]
\begin{center}
\hspace{-1.45cm} (a) \hspace{4cm} (b)
\includegraphics[scale=0.4]{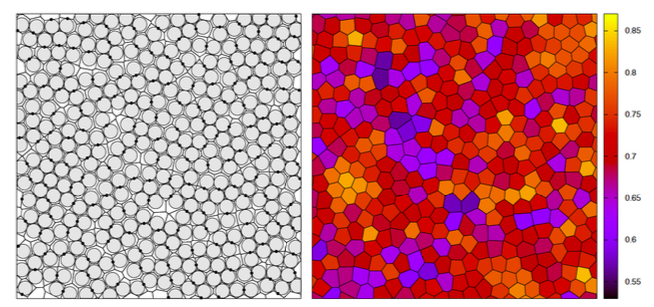}
\\
\hspace{-1.45cm} (c) \hspace{4cm} (d)
\includegraphics[scale=0.4]{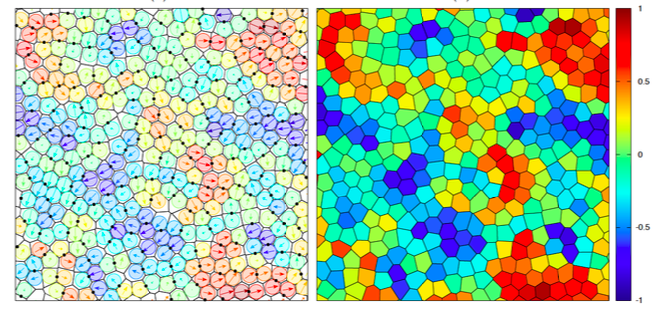}
\end{center}
\caption{(a) The Voronoi tessellation of a dumbbell configuration (the joining point between the two beads is shown with a dark spot). (b) The heat map attributed to the 
local density of the individual disks, with their own Voronoi cells painted accordingly.
In panels (c) and (d) the cells are painted following the heat map attributed to the projection of the 
local hexatic order parameter in the direction of its mean value. In (c) the arrows are still drawn and light 
colours are used while in (d) the arrows have been erased and darker colours are chosen.
}
\label{fig:tessellation}
\end{figure}

A visual inspection of the density fluctuations in real space is achieved by  
painting with the colour that corresponds to its coarse-grained local density each Voronoi region. Typically,  
a heat map with the usual convention, denser in red and looser in blue, is used.
The dumbbell configuration in Fig.~\ref{fig:tessellation}~(a)  is analysed along these lines and the 
local densities are painted according to such a colour code in panel~(b).

\subsubsection{Positional order}

The fluctuating local density field is defined as 
\begin{equation}
\rho({\bf r},t) = N^{-1} \sum_{i=1}^N \delta({\bf r}-{\bf r}_i(t))
\; , 
\end{equation}
where the sum runs over particle indices, ${\bf r}_i(t)$ is the time-dependent position of the $i$th particle, 
and $N$ is the total number of particles. The quantity is normalised to one $\int d^d{\bf r} \, \rho({\bf r},t) = 1$.
In a homogeneous system $\rho({\bf r},t)=\rho_0=1/V$ with $V\equiv\int d^d{\bf r}$ the volume of the box where the 
system is confined.

The density-density correlation function
\begin{equation}
C({\bf r}_1, {\bf r}_2,t) = \langle \rho({\bf r}_1, t)\rho({\bf r}_2, t)\rangle
\end{equation}
involves an average over different realisations of the system henceforth represented by the symbol $\langle \dots \rangle$.
This quantity is, for a homogeneous system, space translational invariant meaning that it only depends on ${\bf r}_1-{\bf r}_2={\bf r}$.
If, moreover, the density is isotropic, $C({\bf r}_1, {\bf r}_2,t) = C(|{\bf r}_1 - {\bf r}_2|,t)$.
 
 The Fourier transform of the density-density correlation yields the {\it structure factor}
 \begin{equation}
 S({\bf q}, t) = \int d^d{\bf r}_1 \int d^d{\bf r}_2 \ NC({\bf r}_1, {\bf r}_2, t) \ e^{-{\rm i} {\bf q} \cdot ({\bf r}_1- {\bf r}_2)}
 = \frac{1}{N} \sum_{ij} \langle e^{-{\rm i} {\bf q} \cdot ({\bf r}_i(t)- {\bf r}_j(t))} \rangle
 \; . 
 \label{eq:structure-factor}
 \end{equation}
 
A real order parameter related to possible translational order is the modulus of the Fourier transform of the density
\begin{equation}
\psi_T({\bf q},t) = \left| N^{-1} \sum_{i=1}^N e^{{\rm i} {\bf q} \cdot {\bf r}_i(t)}\right|^2
\; . 
\label{eq:psiT}
\end{equation}

The positional order can be tested with correlations of the Fourier transform of the density
\begin{equation}
C_{\bold{q_0}} (r,t) = \langle e^{{\rm i} \bold{q_0} \cdot ({\bold r}_i(t) - {\bold r}_j(t))}\rangle
\label{eq:position-correlation}
\end{equation}
where $r = |{\bf r}_i(t) - {\bf r}_j(t)|$ and  $\bold{q}$ is taken to be $\bold{q_0}$
the wave vector  that corresponds to the maximum value of the first diffraction peak of the structure factor (\ref{eq:structure-factor}).
The quasi-long range positional order in the solid phase should be evidenced by an algebraic decay of $C_{\bold{q_0}} (r)$
with distance. Instead,  in the hexatic and liquid phases the decay of $C_{\bold{q_0}} (r)$ should be exponential, see 
Fig.~\ref{fig:sketch-corr}.

\subsubsection{Orientational order}

The  order parameter for orientational order  is 
the local time-dependent $n$-fold order parameter
\begin{equation}
{\psi_n}_i(t) =  \frac{1}{N_{\rm nn}^{i}} \sum_{k(i)}^{N_{\rm nn}^i}  e^{{\rm i} n \theta_{ki}(t)}
\; , 
\label{eq:local-hexatic-op}
\end{equation}
where $\theta_{ki}(t)$ is the instantaneous angle formed by the bond that links the $i$th bead with the $k$th one, the latter being a 
first neighbour on the Voronoi tessellation of space, and a chosen reference axis, see Fig.~\ref{fig:voronoi-hexagonal}~(c). 

The ground state of a system of spherically symmetry constituents interacting {\it via} Lennard-Jones-like potentials is 
expected to be a perfect hexagonal lattice; for a recent review 
on the search for ground state configurations, written from a mathematical 
perspective, see Blanc \& Lewin~\citeyear{Blanc}.  For beads regularly placed on the vertices of such a 
lattice, see  Fig.~\ref{fig:voronoi-hexagonal}~(b), each site has six nearest-neighbours and all the angles are, $\theta_{ki} = 2\pi k/6$. The relevant 
value of $n$ to use is $6$, for which ${\psi_6}_i =1$ on a perfect hexagonal lattice.
At finite temperature this parameter cannot be equal to one.

A colour code is useful to detect local orientational order. The plane is first divided in cells according to the Voronoi prescription, see 
Figs.~\ref{fig:voronoi-hexagonal}~(a)~and~\ref{fig:tessellation}~(a).
Next, an arrow representing the local (complex) order parameter is attributed to each cell, see 
Fig.~\ref{fig:tessellation}~(c). The average over all cells is computed 
yielding a vector of given modulus and direction. The local arrows are then projected on the direction of the averaged vector. The 
cells with maximal projection are coloured in red, the cells with maximal projection in the opposite direction are coloured in blue, and the 
scale in between is colour coded as in the right bar in Fig.~\ref{fig:tessellation} (d).

Upon coarse-graining over a cell with radius $R$ around a point ${\bf r}$ 
 that possibly contains a large number of particles, say $n_R$, one 
can build histograms and then probability distribution functions 
of the coarse-grained local hexatic order parameter 
\begin{equation}
{\psi^{\rm cg}_n}({\bf r},t) = \frac{1}{n_R} \sum_{i\in S_{\bf r}} \frac{1}{N_{\rm nn}^{i}} \sum_{k(i)} e^{{\rm i} n  \theta_{ki}(t)}
\; . 
\end{equation}
With the complex order parameter (\ref{eq:local-hexatic-op}), or with its coarse-grained version, one then constructs a real correlation function
\begin{equation}
C_n(r,t) = \frac{ \langle {\psi^*_{n}}_i(t){\psi_{n}}_j(t)\rangle_{r=|{\bf r}_i-{\bf r}_j|}}{\langle {\psi^*_{n}}_i(t){\psi_{n}}_i(t)\rangle} 
\end{equation}
where homogeneity and isotropy have been assumed on the left-hand-side, with the simple dependence on $r=|{\bf r}_i-{\bf r}_j|$.

The hexatic order parameter and its correlations probe the bond-orientational order thus 
allowing the liquid, hexatic and solid phases to be distinguished: the decay should be exponential in the liquid, 
power law in the hexatic and approach a constant in the solid, see Fig.~\ref{fig:sketch-corr}.

\subsubsection{Defects}

\begin{figure}
\begin{center}
\includegraphics[scale=0.5]{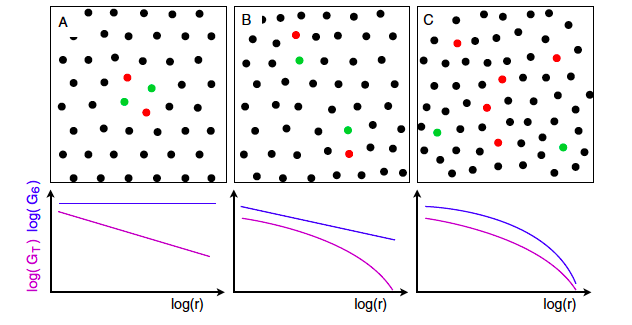}
\end{center}
\caption{Sketch of  configurations with defects of various kinds (first line) and 
the correlation functions (second line) in the solid, hexatic and liquid phases. Image taken from~Gasser (2009).
Red (green) points are particles with five (seven) neighbours. The sketch in 
A is in the solid phase, the one in B in the hexatic phase and the one in C 
in the liquid phase.}
\label{fig:sketch-corr}
\end{figure}

From the Voronoi tessellation of space one can count the number of neighbours that any particle has.
Deviations from the value 6, the one that a perfect hexagonal lattice would have, as in Fig.~\ref{fig:voronoi-hexagonal}~(b), 
are defects, typically corresponding to particles having 5 or 7 neighbours. The sketches in  Fig.~\ref{fig:sketch-corr} show the behaviour 
of the defects, according to the KTHNY scenario: the 
defects are bound and can freely move without destroying the quasi long-range translational and long-range
orientational order  in the solid (A), they unbind in pairs in the hexatic (B) and they get free in the liquid (C).
In actual fact, recent simulations~\cite{Qi14} show that 
defects get together in clusters, which tend to be small and compact in the hexatic phase, 
but become string-like (grain boundaries) in the coexistence region and the liquid phase. 
This is one of the issues that deserves a better analysis in the
future, both for passive and active systems.

\subsubsection{Time delayed correlation functions}

The use of time correlation functions in experiments, instead of spatial ones, has the advantage that due 
to the limited field of view, the latter cannot be obtained over a large dynamic range, 
while the former can, in principle, cover arbitrarily many decades. Time-delayed correlation functions 
simply compare the value of an observable at a time $t_1$ with the value of the same observable at a 
later time $t_2$. The result is a function of $t_2-t_1$ under stationary conditions and the exponential 
or algebraic decays of spatial correlations as a function of distance in the various phases is translated into exponential 
or algebraic decays as functions of time-delay in the temporal ones.

\subsubsection{Mean-square positional displacement}

The simplest time-delayed observable that  characterises the translational properties 
of the elements in the samples is  the averaged mean-square displacement of the position
of a relevant point on the molecules, be it the centre of the disks or the centre of mass,
\begin{equation}
\Delta_r(t-t_w) = \frac{1}{dN} \sum_i \langle |{\bf r}_i(t) - {\bf r}_i(t_w)|^2 \rangle
\; . 
\label{eq:mean-square-disp}
\end{equation}

\subsubsection{Angular mean-square displacement}

Another characterisation of the global dynamics, that focuses on the rotational properties of the constituents,  is given by the averaged mean-square displacement of the angles
formed by a relevant direction of the molecule and a chosen axis of reference, 
\begin{equation}
\Delta_\vartheta(t-t_w) = \frac{1}{N} \sum_i \langle |\vartheta_i(t) - \vartheta_i(t_w)|^2 \rangle
\; . 
\label{eq:mean-square-disp-angle}
\end{equation}

\subsubsection{The Lindemann criterium}

The dynamic Lindemann parameter~\cite{Zahn99} is
\begin{equation}
\gamma_L(t) =   \langle \{ [{\bf r}_i(t)-{\bf r}_i(0)] - [{\bf r}_j(t)-{\bf r}_j(0)] \}^2\rangle /(2a^2)
\end{equation}
with $i$ and $j$ first neighbour particles at the initial time $t=0$, say, and $a$ the crystalline lattice spacing. 
In a crystal $\gamma_L(t) $ is bounded at long times while in the liquid the displacements of the two 
particles are uncorrelated and $\gamma_L(t) = \langle \{ [{\bf r}_i(t)-{\bf r}_i(0)]^2 \rangle/(2a^2) + 
\langle[{\bf r}_j(t)-{\bf r}_j(0)]^2 \rangle /(2a^2)$ is just the mean-square displacement of an individual particle, 
see Eqn.~(\ref{eq:mean-square-disp}), and it is then proportional to $t$ (simple diffusion).
The change from bounded behaviour to time-increasing one can be used as evidence for the end of the solid phase. 

\section{A reminder on phase transitions}

This Section contains  a (very) rapid reminder of phase transitions with emphasis on
properties and concepts that are useful for our purposes. Many textbooks give detailed 
descriptions of critical phenomena~\cite{Stanley71,Amit84,Parisi88,Goldenfeld92,Cardy96,Simon97,Herbut06,Kardar07}.

\subsection{First  order}

In a first-order phase transition a state that is stable on one side of the transition, becomes metastable on the 
other side of it. The order parameter jumps at the transition, for example, from zero in the disordered phase 
to a non-vanishing value in the ordered one. The correlation length, that is extracted from the correlations 
of the fluctuations of the order parameter with respect to its average, is always finite.

In common discussions of this kind of transition, the interplay between only two states is considered, 
each one being the preferred one on the two sides of the 
transition. But this is not necessarily the case and a competition between various equivalent stable states can also 
arise. The dynamics of first order phase transitions
is driven by nucleation of the new stable phase within the metastable one in which the system
is placed initially. During a long period of time the system attempts to nucleate one or more bubbles
of the stable phase until some of them reach the critical size and then quickly grow. In the multi-nucleation problem, 
two possibilities then arise: either one 
of them rapidly conquers the full sample or many of them touch,  get stuck, and a new coarsening 
process establishes. The latter case is the one that will be of interest in the hexatic-liquid transition, as we will 
argue below.

 \subsection{Second order}

In a second-order phase transition a state that is stable on one side of the transition, becomes unstable on the 
other side of it and, typically, divides continuously into an even number of different stable points, related
in pairs by symmetry. The order parameter is continuous at the transition and, for example, it grows from zero 
in the ordered phase. The correlation length, also extracted from the correlations 
of the fluctuations of the order parameter with respect to its average, diverges algebraically on both sides of the 
transition.

When the parameters are taken across the critical value, the system needs to accommodate to the new conditions 
and it does progressively, by locally ordering domains of each of the possible and equivalent new equilibrium states.
The latter process is called coarsening or domain growth and, although it is a very general phenomenon, its details
depend on some characteristics of the problem as the conservation laws and the dimension of the 
order parameter. The symmetry breaking process, whereby one of the equivalent equilibrium states 
conquers the full sample, is achieved late after the system is taken across the phase transition.
Indeed, equilibration takes a time that scales with the system size and diverges in the thermodynamic limit.

\subsection{Infinite order}
 
Berezinskii-Kosterlitz-Thouless (BKT) phase transitions~\cite{BerezinskiI1971a,BerezinskiI1971b,Kosterlitz1972,Kosterlitz1973,Kosterlitz1974,Kosterlitz2016} 
lack an order parameter taking a non-vanishing value on
one side of the transition (in the thermodynamic limit) and are not related to spontaneous symmetry breaking. 
They are transitions of a different kind, driven by the unbinding of 
topological defects when a critical value of a control parameter (typically temperature over an energy scale) is 
reached. In the disordered phase the density of free topological defects is finite and 
the correlation function of the would-be order parameter decays exponentially, with a 
correlation length that is proportional to the distance between unbound defects. This 
length diverges exponentially at the transition and remains infinite in the full quasi-long-range ordered
phase. Topological defects exist in the ordered phase but they bound in pairs and such 
localised in space. The divergence of the correlation length implies that the correlations of the would-be order parameter decay algebraically beyond the transition, that the system has quasi long range order 
and that this full phase behaves as a critical point. In terms of the associated susceptibility, it is finite in the disordered 
phase and it diverges in the full subcritical phase.
Even more so, the transition is characterized by essential singularities in all thermodynamic functions.
The reason for this behaviour are (spin or density) wave excitations with a linear dispersion relation at long wave-lengths.
 
 The dynamics of such phase transitions is characterised by the growth of the quasi-long-range order and the 
 annihilation of topological defects, see, {\it e.g.}~Jeli\'c \& Cugliandolo~\citeyear{Jelic11} for a numerical study of the $2d$XY model
 or Comaron {\it et al.}~\citeyear{Proukakis18} for a similar analysis of a driven-dissipative bidimensional quantum system.
 
 \subsection{From infinite to first order}

The picture described above has been developed based on the  analysis of the $2d$ XY model in which planar spins 
are placed on the vertices of a regular lattice, with nearest neighbour pairwise interactions
$-J {\bf s}_i \cdot {\bf s}_j = -J \cos\theta_{ij}$ with $\theta_{ij}$ the angle between the two spins. 
Interestingly enough, the nature of the transition can change dramatically 
if the interaction term takes  other forms that still respect rotational invariance. 
The potential $2 [1-\cos^{2p^2}(\theta_{ij}/2)]$, that interpolates between the conventional one for $p=1$ and a 
much steeper well for large $p^2$, was used by Domany, Schick and Swendsen~\citeyear{Domany84} to show that the 
transition crosses over from BKT to first order for large  $p^2$, see also Jonsson, Minnhagen and Nylen~\citeyear{Minnhagen93} 
and Zukovic and Kalagov~\citeyear{Zucovic17} for a
more recent and complete analysis of the Langevin and Monte Carlo studies of the same model, respectively. In particular, 
for $p^2 = 50$ the transition is very sharp with a huge peak in the specific heat and many other elements of a first order phase transition. The reason for this behaviour is that 
the typical temperature for the unbinding of vortex-antivortex pairs is pushed to very high values, beyond the ones at which other kinds of excitations
drive the discontinuous transition. Similarly, other examples of models expected to have 
BKT transitions, such as the $2d$ Coulomb gas, were shown to comply with the expectations only at low density
and depart towards a first order phase transition at higher density~\cite{Minnhagen87}. 

\subsection{Two results against common wisdom}

\subsubsection{Breakdown of universality}

The phenomena described by the end of the last Subsection 
seem to be in contradiction with the picture that emerges from the renormalization group theory 
according to which systems in the same universal class (having the same symmetry of the order parameter and 
same dimensionality) should exhibit the same type of phase 
transition with identical values of critical 
exponents. However, a rigorous proof that 
planar spin models of the XY kind with a sufficiently narrow potential 
undergo first order phase transitions was provided by van Enter \& Shlosman 
(2002)~and 
the fact that with a simple change of parameter one can change the order of the transition was thus confirmed.

\subsubsection{Continuous symmetry breaking in two dimensions}

It is commonly found in the literature that the content of the Mermin-Wagner theorem is that
``two-dimensional systems with a continuous symmetry cannot have a broken symmetry at finite temperature''. This, however, is not
true in general and it was not claimed by these authors either. The Mermin-Wagner 
result is that  the expected order parameters vanish in a number of two dimensional cases
cases, including superfluids, superconductors, magnets, and 
crystals, but it does not imply that there cannot be long-range order of another kind. 
The spontaneous breaking of orientational order in $2d$, that was already anticipated by Landau 
himself~\cite{Landau37a,Landau37b}, is a counter example of the statement between inverted commas.
Indeed, broken translational symmetry implies broken rotational symmetry. However, 
the converse is not true and it is possible to
break rotational invariance without breaking translational invariance. The most obvious way to do it is to 
use anisotropic molecules, as in liquid crystal systems. Another way is by achieving long-range {\it bond orientational}
order with spherically symmetric constituents.

\section{Equilibrium phases in two dimensions}

In this Section we recall some properties of passive matter in equilibrium. We focus on two  representative cases.
We take a dense system at low temperatures and we evaluate the effect of thermal fluctuations.
We consider a lose system and we derive the virial expansion that leads to the equation of state. These 
two well-known problems have intriguing extensions when activity is added, issues that we will 
cover in the next Section. 

\subsection{Effects of fluctuations on dense systems: melting in low dimensions}

Consider a sufficiently dense system so that it should be a solid, possibly in a crystalline phase
and evaluate the effect of thermal fluctuations. Does the solid melt? Which are the mechanisms
leading to melting? Which is the order of the phase transition taking the solid into a liquid?

\subsubsection{Positional vs. orientational order}

In the 30s \shortciteN{Peierls34} 
and Landau
\citeyear{Landau37a,Landau37b}
argued that it is not possible to find long-range positional 
order in low dimensional systems with short-range interactions. 

Peierls used the simplest possible model for a solid, one of beads placed on a $d$-dimensional lattice, with Hookean couplings
between nearest-neighbours, in canonical equilibrium. The question he asked was whether such a system could 
sustain periodic order over long distances under thermal fluctuations, and he concluded that this is not possible 
in $d\leq 2$, while it is in $d\geq 3$. Landau based his arguments instead on his theory of phase transitions and reached the 
same conclusion.
In the 60s, the numerical simulations of~\shortciteN{Alder1962} pointed towards a first order phase transition 
between solid and liquid.
A more general proof of absence of crystalline order in $2d$, that does not rely on the harmonic approximation but uses 
a classical limit of Bogoliubov's inequality~\cite{Bogoliubov62}, was given later by~\shortciteN{Mermin68}.

An equilibrium amorphous state has a uniform averaged density $\langle \rho\rangle =\rho_0$,
while a zero temperature crystalline state has a periodic one
\begin{equation}
\rho({\bf r}) = \sum_{i} \delta({\bf r}-{\bf R}_i)
\end{equation}
with 
$i$ a label that identifies the particles or lattice sites,  and ${\bf R}_i$ 
the position of the $i$th vertex of the lattice. At zero temperature a perfectly ordered state, 
with  periodic density is allowed for all $d\geq 1$.
However, thermal fluctuations make the atoms vibrate around their putative lattice sites, 
and the instantaneous position of the $i$th atom becomes 
\begin{equation}
{\bf r}_i = {\bf R}_i+{\bf u}_i = {\bf R}_i+{\bf u}({\bf R}_i)
\end{equation}
with ${\bf u}_i={\bf u}({\bf R}_i)$ its displacement from ${\bf R}_i$. 
A simple way to see the lack of positional order in low dimensions (and the existence of it in higher dimensions) 
is to compute the mean-square displacement of the atoms assuming thermal equilibrium. Take a 
generic pair-wise potential
\begin{equation}
U_{\rm tot} = \frac{1}{2} \sum_{ij} U({\bf r}_i - {\bf r}_j) = \frac{1}{2} \sum_{ij} U({\bf R}_i - {\bf R}_j + {\bf u}_i - {\bf u}_j)
\; . 
\end{equation}
Indeed, 
the total harmonic potential energy is~\cite{AshcroftMermin}
\begin{eqnarray}
U_{\rm tot} &=&
U_{\rm gs} + 
\frac{1}{2} \sum_{ij} \sum_{\mu\nu} 
(u^\mu_i-u^\mu_j) \; \frac{\partial^2 U}{\partial_{r^\mu_i}\partial_{r^\nu_j}}({\bf R}_i-{\bf R}_j)  \;  (u^\nu_i-u^\nu_j)
\nonumber\\
&=& 
U_{\rm gs} + 
\frac{1}{2} \sum_{ij} \sum_{\mu\nu} 
u^\mu_i D_{\mu\nu}({\bf R}_i-{\bf R}_j)  u^\nu_j
\end{eqnarray}
where $U_{\rm gs} = \frac{1}{2} \sum_{i\neq j} U({\bf R}_i - {\bf R}_j)$, 
and in the second term
$\mu, \nu$ run from 1 to $d$,
$D^{\mu\nu}_{ij} \equiv D_{\mu\nu}({\bf R}_i-{\bf R}_j) = \delta_{ij} \sum_k \phi^{\mu\nu}_{ik}- \phi^{\mu\nu}_{ij}$
and
$\phi^{\mu\nu}_{ik} = \partial^2 U({\bf r})/\partial r^\mu_i \partial r^\nu_k$. 
Three symmetries of the couplings follow immediately
$D^{\mu\nu}_{ij}=D^{\nu\mu}_{ji}$,
$D^{\mu\nu}_{ij}=D^{\mu\nu}_{ji}$ 
(from the inversion symmetry of a Bravais lattice), 
and 
$\sum_i D^{\mu\nu}_{ij} = 0$
(from the uniform translation invariance of the full lattice).
After a Fourier transform $U_{\rm tot}$ becomes
\begin{eqnarray}
U_{\rm tot} = U_{\rm gs} + 
 \frac{1}{2} \sum_{\bf k} \sum_{\mu\nu}
\tilde u_\mu^*({\bf k}) \tilde D_{\mu\nu}({\bf k}) \tilde u_\nu({\bf k})
\; , 
\end{eqnarray}
where 
$\tilde u_\mu({\bf k}) =  \sum_i e^{{\rm i} {\bf k} \cdot {\bf r}_i} {\bf u}_i$
and $\tilde u^*_\mu({\bf k}) = \tilde u_\mu(-{\bf k})$ since 
${\bf u}_i$ is real.
Next one needs to estimate the ${\bf k}$ dependence of 
$\tilde D_{\mu\nu}({\bf k})$. Using the symmetries of 
$D^{\mu\nu}_{ij}$, its Fourier transform
$\tilde D_{\mu\nu}({\bf k})$ can be recast as 
\begin{equation}
\tilde D_{\mu\nu}({\bf k}) = -2 \sum_{{\bf R}} D_{\mu\nu}({\bf R}) \sin^2 ({\bf k} \cdot {\bf R}/2)
\approx -2 \sum_{{\bf R}} D_{\mu\nu}({\bf R}) ({\bf k} \cdot {\bf R}/2 )^2
\; , 
\end{equation}
after a small ${\bf k}$ approximation.  It is now possible to further assume 
\begin{equation}
\tilde D_{\mu\nu}({\bf k}) \mapsto k^2 A_{\mu\nu}
\end{equation}
where the important $k^2$ dependence has been extracted. 
$U_{\rm tot}$ thus becomes the energy of an ensemble of 
harmonic oscillators. 
The equipartition of quadratic degrees of freedom in canonical equilibrium yields
\begin{equation}
\langle \tilde u^*_\mu({\bf k})\tilde u_\nu({\bf k})\rangle
 = \frac{k_BT}{k^2}\ A^{-1}_{\mu\nu}
\end{equation}
and a logarithmic divergence of the displacement mean-square displacement
\begin{equation}
\Delta u^2 \equiv \langle |{\bf u}({\bf r}) - {\bf u}({\bf r}')|^2 \rangle 
\sim k_BT \ln |{\bf r} - {\bf r}'| \qquad\mbox{in} \;\; d=2
\label{eq:msd-u}
\end{equation}
 follows as a consequence of the logarithmic divergence of the integral $\int d^2k \ k^{-2}$. 

An even simpler derivation of the same result goes as follows. 
Take the harmonic Hamiltonian $H = \frac{c}{2} \int d^d{\bf r} \; ({\bf \nabla} {\bf u})^2$
as a starting point. The excitation of a spin-wave with wavelength $L$ (wave vector $2\pi/L$)
then requires an energy $E\approx L^d (2\pi/L)^2 \propto L^{d-2}$
that diverges with $L$ for $d=3$, is independent of $L$ for $d=2$  (marginal case)
and  decreases as $L^{-1}$ for $d=1$.

The divergence of the mean-square displacement  in Eqn.~(\ref{eq:msd-u}) 
implies that any atom displaces a long distance from each other 
and hence no long-range order is possible in $d=2$. 
This weird effect is due to the dimensionality of space. 
In three dimensions, the mean square fluctuation is finite.

A more general proof of the lack of positional order in $d\leq 2$ that goes beyond the 
harmonic approximation was by Mermin~\citeyear{Mermin68}. In this paper, 
he first proposed the following criterium for crystallinity:
\begin{eqnarray}
\begin{array}{rcl}
\tilde \rho({\bf k}) &=& 0 \qquad \mbox{for} \;\; {\bf k} \;\; \mbox{not a reciprocal lattice vector}
\; , 
\\
\tilde \rho({\bf k}) &\neq & 0 \qquad \mbox{for at least one non-zero reciprocal lattice vector}
\; , 
\end{array}
\label{eq:condition}
\end{eqnarray}
with $\tilde \rho({\bf k})$ the Fourier transform of $\rho({\bf r})$, in the thermodynamic limit, that is
\begin{equation}
\tilde \rho({\bf k}) = \frac{1}{N} \sum_{i=1}^N e^{{\rm i} {\bf k} \cdot {\bf r}_i}
\; . 
\end{equation}
Using Bogolyubov's identity, Mermin showed that the condition (\ref{eq:condition}) 
cannot be satisfied in $d\leq 2$ since in thermal equilibrium at non-vanishing temperature,
for all ${\bf k}$, $\langle \tilde \rho({\bf k}) \rangle$ 
is bounded form above by a quantity that vanishes in the thermodynamic limit.

The possibility of a two-dimensional system with constant density (all Fourier modes vanish) being, however, 
anisotropic over long distances was left open by Peierls and Landau. The actual definition of the orientational order was 
also given by Mermin in his 1968  paper. Within the harmonic solid model he simply noticed that 
\begin{equation}
\langle [{\bf r}({\bf R}+{\bf a}_1) - {\bf r}({\bf R})] \cdot   [{\bf r}({\bf R}'+{\bf a}_1) - {\bf r}({\bf R}')] \rangle 
\end{equation}
approaches $a_1^2$ at long distances $|{\bf R}-{\bf R'}| \to \infty$, implying that  the orientation of the local 
order is maintained all along the sample.
The status of the studies of orientational order in two dimensional systems in the 90s is summarised 
in~\cite{Strandburg92}.

\subsubsection{Melting scenarii}
In $d\geq 3$ melting is a first order phase transition between crystal and 
liquid (although the details of how this transition occurs are still not fully understood and may depend on the 
material). In $d=2$, instead,
there is no full consensus yet as to which are the mechanisms for melting and 
how the passage from solid (with quasi-long-range positional and long-range orientational order) to liquid (with 
both short-range positional and orientational order) occurs. In the late 70s Halperin \& Nelson~\citeyear{Nelson1979}
and Young~\citeyear{Young1979} suggested that the transition can occur in two steps, with an intermediate 
anisotropic {\it hexatic} phase with short-range positional and quasi-long-range orientational order. Both transitions, between solid and 
hexatic on the one hand, and hexatic and liquid on the other, were proposed to be driven by the dissociation of topological 
defects, and therefore be of BKT type:
\begin{itemize}
\item
 In the first stage, at  the melting transition $T_m$, dislocation pairs unbind to form a bond orientationally ordered hexatic liquid. 
 \item
 In the second stage, at $T_i$, 
 the disclination pairs which make up the dislocations unbind to form an isotropic liquid. 
 \end{itemize}
 Moreover,  whithin the KTHNY theory, the {\it finite size scaling} of the order parameters is expected to be as follows.
In the solid phase the translational order parameter should decay with system size as $N^{-\eta}$ with $\eta\leq 1/3$.
In the hexatic phase  the hexatic order parameter should decay with system size as $N^{-\eta_6}$ with $\eta_6 \to 0$
at the transition with the solid and, according to Nelson \& Halperin, $\eta_6\to 1/4$ at the transition with the 
liquid. All these conclusions were derived from an RG analysis of the continuous elastic model of a solid separated into the 
 contribution of the smooth displacements  and the one of the defects.
 
 A large number of numerical and experimental attempts to confirm (or not) this picture followed. A summary 
 of the situation at the beginning of the 90s can be found in Strandburg (1989, 1992) and 
 close to ten years ago in~Gasser~\citeyear{Gasser09}.
 Early numerics and experiments faced some difficulty in  
establishing the existence of the hexatic phase, and suggested instead  coexistence between solid and 
 liquid as expected in a single first order phase transition scenario. 
 However, by the turn of the century 
 the existence of the hexatic phase was settled and 
 quite widely accepted (see the references by Maret {\it et al.} cited below)
 although evidence for both transitions being of BKT kind remained still elusive.
 
 More recently, Krauth and collaborators~(\shortciteNP{BeKr11,Engel13,KaKr15})
came back to this problem with powerful numerical techniques and they suggested that, for sufficiently hard
repulsive interactions between disks, the transition between the hexatic and liquid phases is of first order. A phenomenon similar to the 
one put forward by~Domany, Schick and Swendsen~\citeyear{Domany84} with a numerical study, and later shown 
rigorously by van Enter and Shlosman~\citeyear{vanEnter02}, would then be at work. Namely, that the BKT transition derived with 
renormalisation group techniques would be preempted by a first order one. This new scenario allows for co-existence of 
the liquid and hexatic phases in a finite region of the phase diagram. The mechanisms for the transitions would then be the following.
\begin{itemize}
\item
 In the first stage, at $T_m$,
 dislocation pairs unbind to form a bond orientationally ordered hexatic phase. 
 \item
 In the second stage, at $T_i$,  grain boundaries made of strings of alternating five and seven fold defects would
 percolate across the sample and liquify it. 
 \end{itemize}

While real time video microscopy on superparamagnetic colloids 
interacting via a {\it soft $r^{-3}$ potential} tend to confirm the KTHNY 
scenario~\cite{Zahn99,Zahn00,vonGrunberg07,Keim07,Gasser10},  
experimental evidence for the new scenario in a colloidal {\it hard disks} system was recently given by 
Thorneywork {\it et al.}~\citeyear{Dullens}. It seems plausible that the mechanism for melting in $2d$ be non-universal and 
depend on the interaction potential and other specificities of the systems. Indeed, the numerical simulations prove that 
for sufficiently soft potential the first order transition is replaced by the conventional BKT one~\cite{KaKr15}. Moreover,  
a choice between the two is also made by the {\it form} of the particles: the polygon case was carefully studied by~Anderson {\it et al.} \citeyear{Glotzer17}
and a dependence of the order of the transition with the number of sides of the 
constituent polygons was claimed in this paper.

\subsection{Effects of interactions on dilute systems: the equation of state}

Since we may be dealing with first order phase transitions, it is useful to recall how these arise in 
the best known case of the liquid-gas transition and how they lead to co-existence. This is seen, for 
instance, from the equation of state derived under various approximations. For example, 
the virial expansion is a common technique used to study weakly interacting gases with perturbative methods.
It is explained in many textbook, see {\it e.g.}~\cite{Kardar07}, and we will not reproduce much details here.
In short, the virial expansion expresses the deviations from  the ideal gas equation of state, $PV =Nk_BT=nRT$ with 
$n=N/N_A$ the number of atoms over Avogadro's number and $R=k_BN_A$ the gas constant, as a power series in the 
density $\rho=N/V$ with temperature dependent coefficients.  Truncated to order $\rho^2$, this expansion 
yields the {\it Van der Waals equation} $P_{\rm eff} V_{\rm eff} = (P+aN^2/V^2) (V-bN) = nRT$ (where the effective 
volume takes into account the reduction due to the space occupied by the particles themselves and the effective 
pressure is higher than the bare one due to the attraction between the particles). 
The latter breaks the positivity requirement on the isothermal compressibility 
$\kappa_T=-V^{-1} \partial V/\partial P|_T$ that is a consequence of its fluctuation-dissipation relation 
(in the macrocanonical ensemble) with 
the variance of the particle number confined to the volume $V$, 
\begin{equation}
 \kappa_T =  \frac{1}{k_BT} \frac{V}{\langle N\rangle^2} \langle (N- \langle N\rangle)^2\rangle  \geq 0
\; . 
\end{equation} 
In physical terms, a system with negative compressibility is unstable and it would collapse. Indeed, 
the Van der Waals isotherms have a portion with negative compressibility that indicates an instability towards 
formation of domains of low and high density, in other words, phase separation between liquid and gas, both with 
positive compressibility. For volumes in this region, the isotherms of the real system are instead flat due to the 
coexistence of the two phases. The {\it Maxwell construction}, an equilibrium argument, 
indicates that the stability of the sample is obtained
at a value of the pressure $P$ (that determines the volumes occupied by the two phases) such that the areas above and 
below the dip and peak of the $P(V)$ curve are the same. 

In a real system with finite size, the isotherms also show a Mayer-Wood loop structure~\cite{MayerWood65}. However, one has to be 
careful before concluding that such a loop is due to a first order phase transition. 
Actually,  finite systems undergoing a 
second order transition may also show one~\cite{Alonso99}. The scaling of the loop area with system size does, instead,
provide unambiguous evidence for first order phase transitions. Indeed, if there is a  coexistence region, the system should hold 
in it an interface between the two macroscopic phases. The surface occupied by this interface should scale as 
$L^{d-1}$ (ignoring possible fractal phenomena) and the free-energy cost of it should therefore be $\Delta F\propto L^{d-1}$
leading to a free-energy density cost of  $\Delta f\propto L^{d-1}/L^d=L^{-1}$ that, in $d=2$ corresponds to 
$N^{-1/2}$. The free-energy density can be derived from the equation of state {\it via} an integration since 
$P = -\partial F/\partial V|_T$ that 
implies that the area occupied by the loop should decrease with system size as $N^{-1/2}$.

\section{Active systems}

We now enter the field of active systems. We very briefly mention in this Section  the results of numerical studies that led us to 
construct the phase diagrams [in the (Pe, $\phi$) plane] of the active Brownian particle and active dumbbell systems.   
When constructing these phase diagrams we seriously took into account the knowledge of the passive limit 
behaviour that we have described so far.

\subsection{Numerical methods}

We will not expose here the numerical methods used to integrate the dynamics of 
passive and active matter system as there exist excellent textbooks and 
review articles in the literature that explain in detail these techniques~\cite{allen}. In a few words,
the integration of Eqns~(\ref{eq:active-BD}) is typically done with the velocity Verlet algorithm~\cite{Rahman64,Verlet67}
that can be easily parallelised, and this is done using the 
LAMMPS method~\cite{plimpton1995fast}.

It is important to note, though, that while in equilibrium 
statistical averages are usually exchanged with time averages under the {\it ergodic hypothesis}, out of 
equilibrium this hypothesis cannot be taken for granted and statistical averages may need to be done
by definition, that is, by averaging over many samples run under the same conditions. 

Numerical simulations suffer from severe limitations given by the typically small size of the 
systems compared to the thermodynamic limit in which theoretical calculations are performed.
Furthermore, the dynamic equations are integrated over relatively short times. In the context of active 
matter, these two limitations can, however, represent the actual experimental situation as 
real systems do not count with constituents as numerous as the Avogadro number 
and time-scales can be relatively short as well. In finite size systems, special care has to be 
taken with the choice of boundary conditions and how these may affect the 
behaviour of the system in the bulk. Periodic boundary conditions are often chosen since they tend to minimise
finite-size effects, not having edges, but they also avoid the annoying decision to make 
concerning interactions between particles and walls. Special care has to be taken not to inhibit periodic or orientational 
order or symmetry with this choice.

Having said this, we are interested in determining phase diagrams that exist in the thermodynamic 
limit. Finite size scaling should therefore be used to determine the behaviour in the infinite size limit.
Typical simulations ran with $N=256^2$ particles, scanning the parameter space  $\phi\in[0:0.9]$ and Pe $\in[0:200]$. 
Notice that since the disks are not completely hard, some overlap between them is possible and 
values of $\phi$ that are slightly larger than the close packing limit can be accessed in the simulation (recall that the close
packing fraction of disks in two dimensions is achieved by a perfect triangular lattice and it amounts to $\phi_{\rm cp} \approx 0.91$).

\begin{figure}[h!]
\hspace{1.7cm} (a) \hspace{5.5cm} (b)
\begin{center}
\includegraphics[scale=0.09]{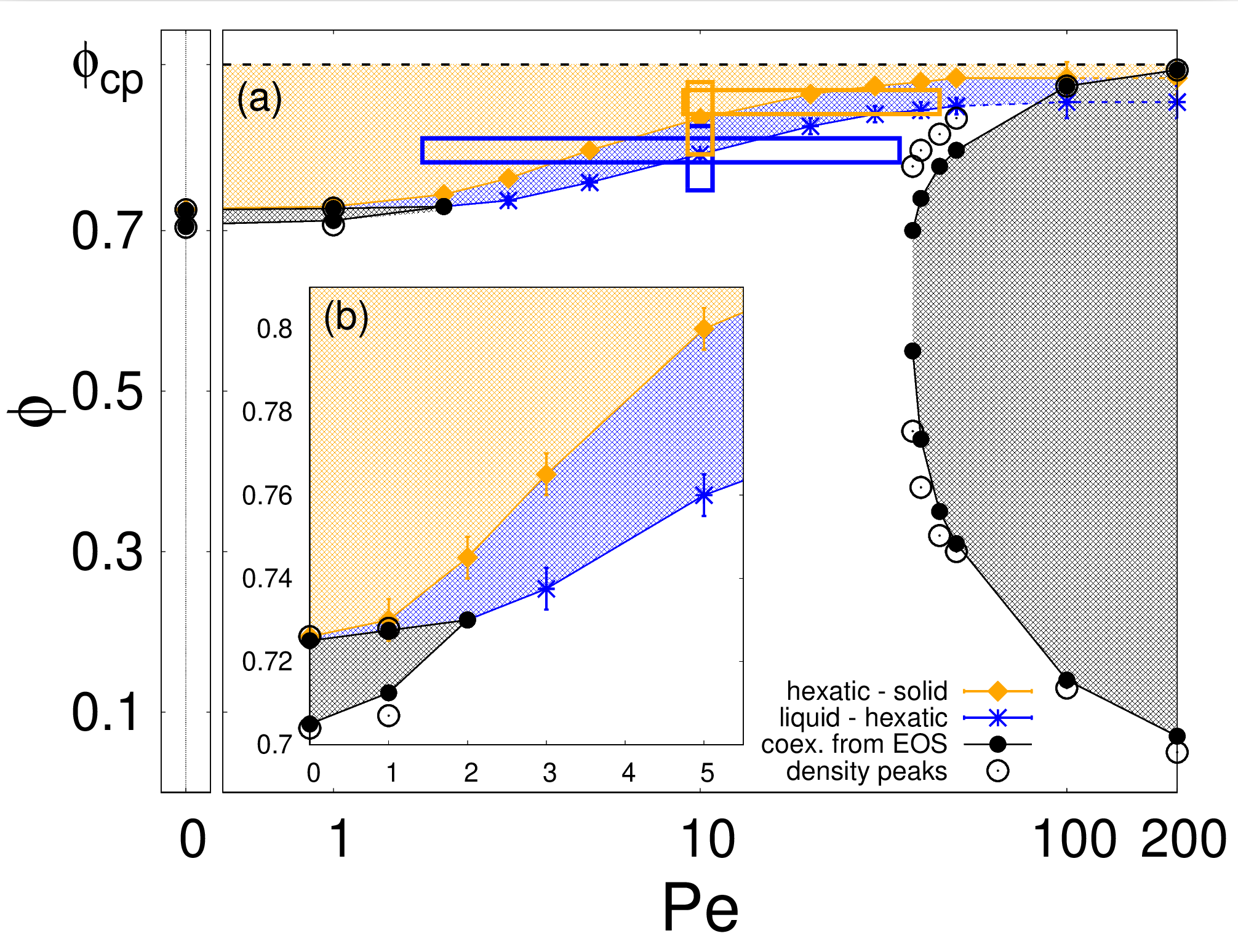}
\includegraphics[scale=0.09]{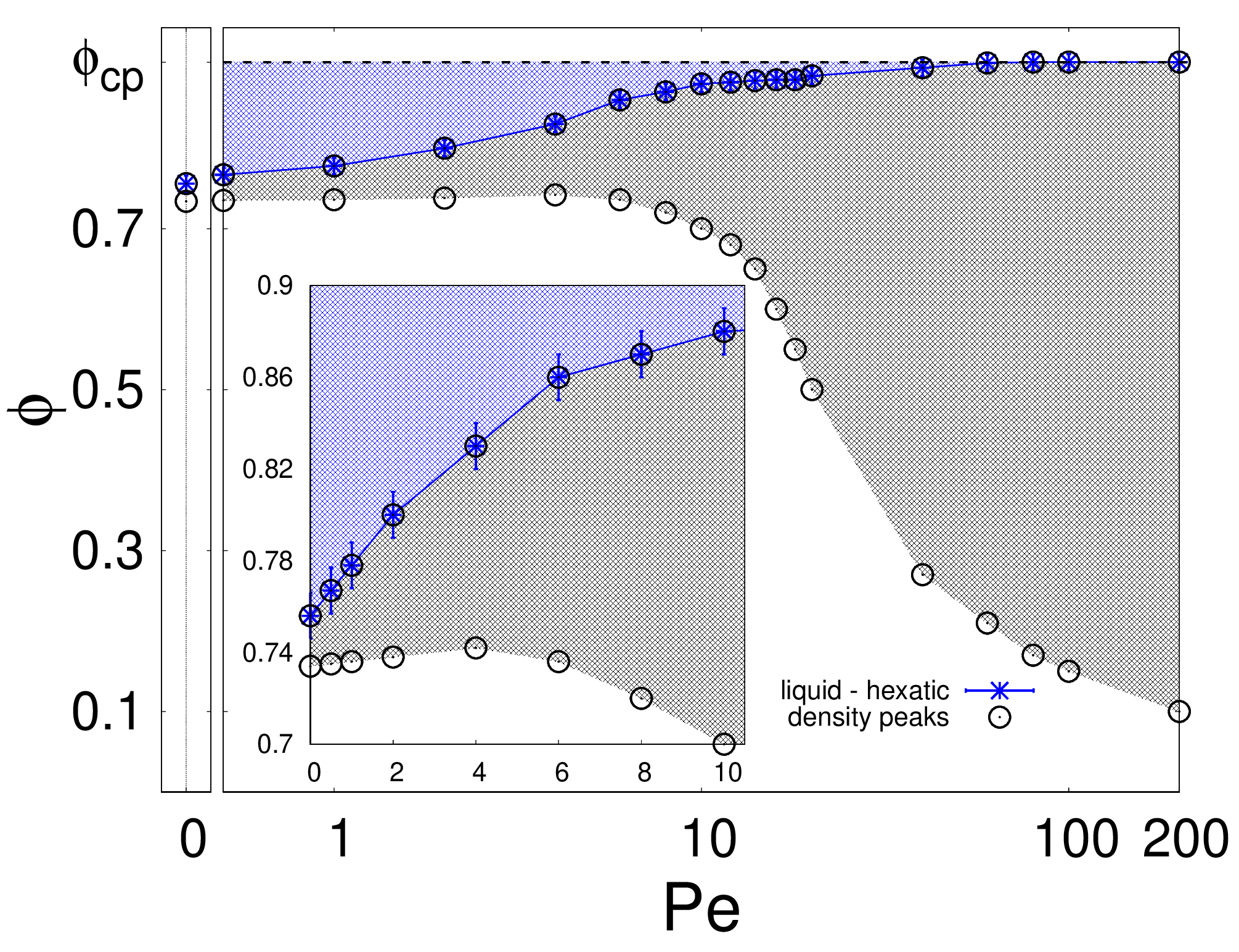}
\end{center}
\caption{The phase diagrams of interacting active Brownian disks (a) and dumbbells (b) as obtained by Digregorio~{\it et al.}~(2018)  for the disks [see also Klamser {\it et al.} (2018)] and Cugliandolo {\it et al.} (2017) and Petrelli {\it et al.} (2018) for the 
dumbbells. The insets are zooms over the small Pe regions, close to $\phi=0.7$.}
\label{fig:phase-diagram}
\end{figure}

\subsection{Phase diagrams}

Passive systems in two dimensions have liquid, orientationally ordered and solid phases.
The fact that there is co-existence between liquid and hexatically ordered phases in systems of passive, 
purely repulsive, disks was evidenced by Bernard, Kampfer \& Krauth, in a series of papers, 
if the potential is hard enough. The question naturally arises as to whether the passive phases, and 
the co-existence region, survive under activity, and whether they do both for spherically symmetric
and elements with non-spherical form.

The various phases can be examined with the usual observables already defined and 
mentioned in the context of the passive limit: order parameters, their correlation 
functions, the distributions of their local values,  their fluctuations, and the pressure 
loop (if accepted, see below).

We will not include here all the evidence for these claims, that can be found in the relevant references, and 
give rise to the two phase diagrams in the (Pe, $\phi$) plane shown in Fig.~\ref{fig:phase-diagram} (a) for the disks and (b) for the dumbbells.  The colour code is such that white is 
liquid (or gas), grey is coexistence, blue is hexatic, and yellow is solid 
(in Sec.~\ref{subsec:phase-sep} we will explain why we have not distinguished hexatic from solid in the dumbbell system and why we have depicted all the region above the end of coexistence in blue).
We will simply discuss in the following paragraphs some aspects of the structure of these systems. 

In both cases there is a region with co-existence (grey) that penetrates the phase diagram along the Pe $\stackrel{>}{\sim} 0$ 
direction, close to  $\phi\simeq 0.7$. The dilute phase has no order and behaves as 
an active liquid or gas while the dense phase next to it has orientational order and it is therefore an active hexatic phase.
However, we see a difference. While for disks this coexistence ends at a relatively small 
value of Pe and the so-called {\it motility induced phase separation} (MIPS) region appears for much larger values of Pe, 
for dumbbells the region with coexistence simply merges (or is the same as) 
what is usually called MIPS. We will not discuss here the transition between the active hexatic (blue) and the active solid (yellow) phases.

\subsection{The equation of state}

The mere definition of pressure needs attention in active matter systems. In mechanical terms, it 
is the force per unit surface exerted by the system on the walls of its container. In equilibrium
in the thermodynamic limit,  the mechanical pressure is also given by an {\it equation of state} that relates it 
to bulk properties of the system, namely, temperature and density, with no reference of the particular interaction potential between constituents
and boundary walls. This relation is not at all obvious out of equilibrium, and it has been observed in some active matter
systems that the pressure does depend on the potential details.  This issue will be covered in other Chapters 
in this book.

Expressions for the mechanical pressure for spherically symmetric constituents
governed by a {\it Markov stochastic process of Langevin kind} both 
under and over damped for confined and periodic boundary conditions, were recently derived. Moreover,
formul\ae $\;$ for the cases of interest here, 
ensembles of active Brownian particles in interaction~\cite{Winkler15}  and of
active dumbbells also in interaction~\cite{Joyeux16} were also recently deduced and discussed.
In cases in which the constituents are not symmetric there is some ambiguity related to the way 
in which the interactions with the walls should be considered; we will not discuss this issue 
further here.

\begin{figure}[t!]
\hspace{1cm} (a) \hspace{5cm} (b)
\begin{center}
\includegraphics[scale=0.5]{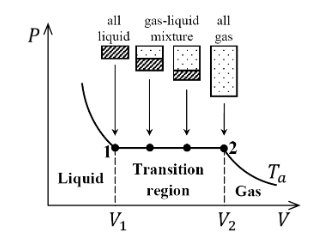}
\includegraphics[scale=0.3]{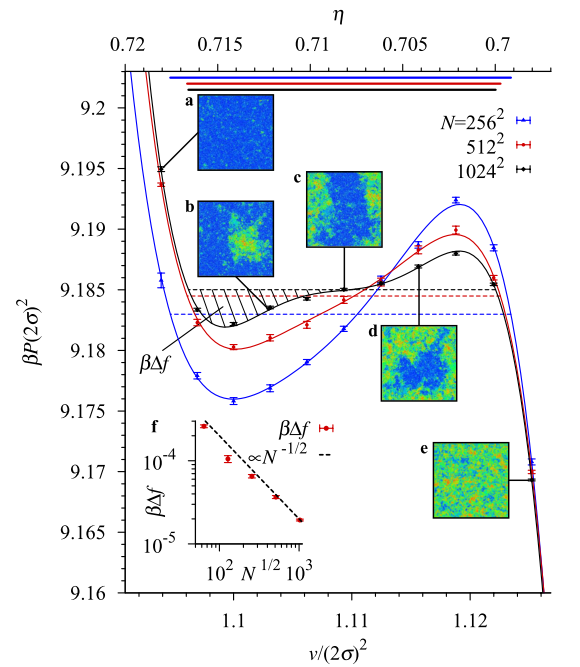}
\end{center}
\hspace{1cm} (c) \hspace{5cm} (d)
\begin{center}
\includegraphics[scale=0.6]{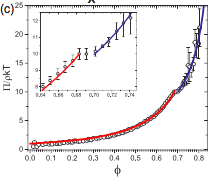}
\hspace{0.25cm}
\includegraphics[scale=0.3]{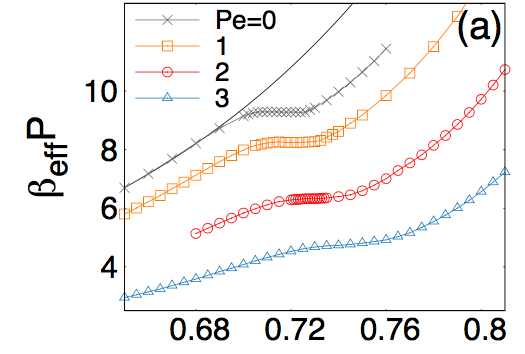}
\end{center}
\caption{The equation of state. (a) The van der Waals equation. (b) Simulation of hard disks from Bernard \& Krauth (2011). 
(c) Experiments on hard disks from Thorneywork {\it et al.} (2017). (d) Simulations of active hard disks from Digregorio 
{\it et al.} (2018). }
\label{fig:eos}
\end{figure}

Figure~\ref{fig:eos} displays four panels with the equation of state of (a) the van der Waals equation, 
(b) the numerical simulations of a rather hard disk passive system from Bernard \& Krauth (2011), (c) the experimental measurements of another 
rather hard disk passive ensemble from Thorneywork {\it et al.} (2017), and (d) the numerical simulations of an ensemble of active 
Brownian disks at small Pe from Digregorio {\it et al.}~\citeyear{Digregorio2018}. All plots are quite similar and were used to claim that the transition is 
of first order in the (b), (c) and (d) cases. In the case of active Brownian particles, these measurements were used to locate 
datapoints that yield the limits of the co-existence region (and coincide with the location of the boundary estimated from the 
distribution of local densities and local hexatic order).

\subsection{Phase separation in  the dumbbell system}
\label{subsec:phase-sep}

We will end this presentation giving some details on the phase separation found in the 
dumbbell system, with molecules made of two joined disks with the same diameter $\sigma_{\rm d}$ and distance between their centres very close to $2\sigma_{\rm d}$ and almost constant over time. A much more 
exhaustive discussion of the structure and 
dynamics of this model can be found in (Petrelli {\it et al.}, 2018).

Take the passive dumbbell system at a given global density with co-existence.
As the activity is turned on, some spatial regions get denser, leaving away disordered holes. Under increasing activity, 
the high density peak in the bimodal distribution of local densities continuously
moves towards higher values  and its weight increases while the low density peak moves in the opposite
direction and its weight decreases.  As far as density is concerned
we do not see any discontinuity when moving towards higher activities in the coexistence region.
A similar behaviour is observed when following the local hexatic order parameter.
Curves of constant repartition of dense and lose phases can be traced and the 
system's behaviour can be compared on these. 
This is different in the active Brownian disk system, for which the region with 
coexistence at low Pe ends on curves at relatively low values of Pe, see the phase diagrams shown in Fig.~\ref{fig:phase-diagram}.

\begin{figure}[h!]
\begin{center}
(a) \hspace{3cm} (b) \hspace{3cm} (c) \hspace{2cm} $\;$
\includegraphics[scale=0.1]{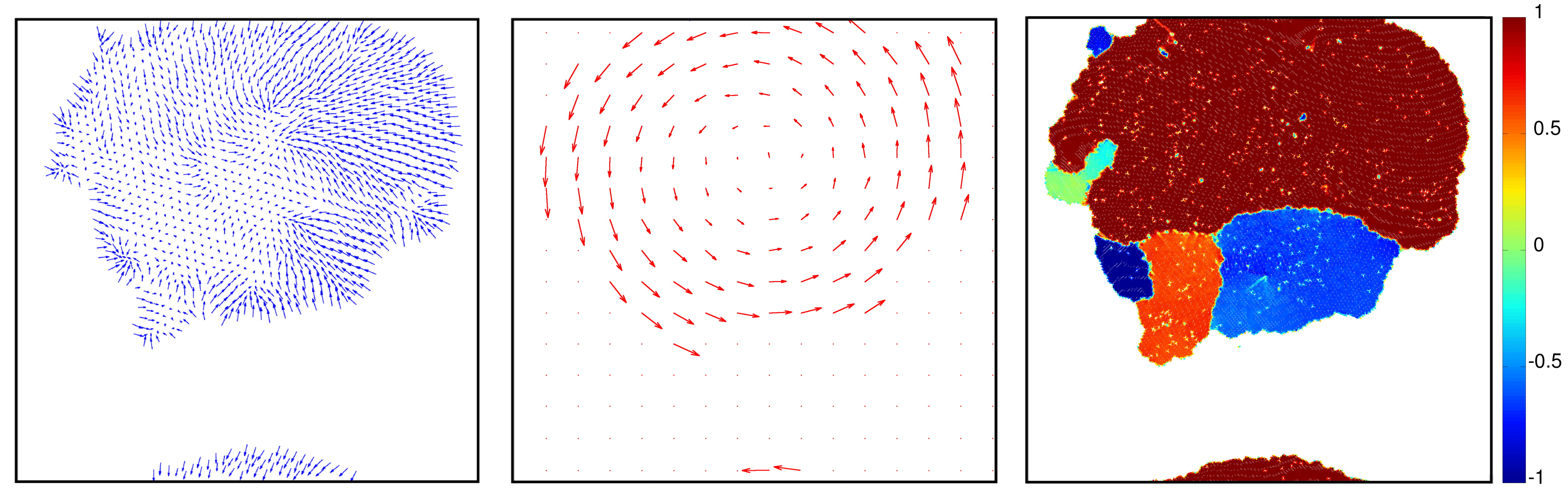}
\end{center}
\caption{Dumbbells at a global packing fraction such that the system is in the coexistence region with 50-50
proportion of liquid and hexatic phases, Pe = 200. Panel (a) shows the dumbbells polarisation, panel (b) the 
local velocity and (c) the local hexatic order parameter. The figure is taken from Petrelli {\it et al.} (2018).}
\label{fig:dumbbells-panels}
\end{figure}

Depending on the strength of the activity, this process allows the dumbbells in the ordered regions 
to pack in a single domain with perfect hexatic order, or in polycrystalline arrangements concerning 
the orientational order, see panel (c) in Fig.~\ref{fig:dumbbells-panels}.  

Another interesting observable to characterise the structure and dynamics of the 
clusters is the coarse-grained {\it polarisation} or, basically, a vector constructed as the average over a coarse-graining volume of the sum of vectors pointing from head-to-tail and attached to each dumbbell, see panel (a) in Fig.~\ref{fig:dumbbells-panels}.
At Pe = 0 there is no polar order whatsoever (not even locally). At intermediate
activity, say Pe = 40, the clusters show an aster polar configuration with a defect at the centre (basically on the 
dumbbell that triggered aggregation and the subsequent formation of the cluster).
At larger activity, say Pe = 200, the aster evolves to a spiralling pattern (see the map in the figure) and the cluster consequently rotates, a motion 
that is not observed at lower Pe, see Fig.~\ref{fig:dumbbells-panels}. We therefore found that polar order
differentiates between clusters at small and high activity. 

Interest can then be set on the motion of the dumbbells.
In the phase separated cases, the dumbbells in the dilute phase are basically free, since their {\it kinetic energy} 
is very close to the one of independent dumbbells for all Pe values. 
The kinetic energy of the dumbbells in the dense phase, instead,  
increases very weakly with activity, due to the fact that the mobility is suppressed 
inside the clusters and these are massive and move very slowly.

While the kinetic energy gives a measure of the strength of 
flow in the system, the {\it enstrophy} is a measure of the presence of vortices in the 
velocity field and it can be used to understand whether clusters rotate
driven by activity.  For high values of the activity the dumbbells
arrange in spirals, the clusters in the system undertake
a rotational motion and the probability distribution
function of the enstrophy develops a multi-peak
structure associated to the rotating clusters. 
A comparison between the velocity and the polarisation fields shows that the 
former  exhibits a vortex pattern while the polarisation one is a spiral.

For strong enough Pe, say Pe $>$ 50, the spontaneously
formed clusters turn around their centre of mass with an
angular velocity that is proportional to the inverse of their
radii. The poly-crystalline nature of the clusters, with respect
to the hexatic order, does not seem to play a major
role in their rotational properties. Instead, the orientation
of the dumbbells inside the clusters is, indeed, important,
as a certain amount of disorder is needed to make them
turn.

The exact nature of the transition between the hexatic and solid phases for the dumbbell
system and its location in the phase diagram are still open questions. 
Since monomers are constrained to be attached in pairs,
they cannot arrange on a triangular lattice at any $\phi<\phi_{\rm cp}$ forcing the 
positional correlations of the monomers
to decay exponentially. It is therefore hard to identify the sold phase and this is the reason why in Fig.~\ref{fig:phase-diagram} (b)  we show all the region of the phase diagram lying above the end of coexistence in blue.


\section{Concluding remarks}

The motivation for the studies of active systems in two dimensions described here was to determine their full phase diagram linking the 
strong activation limit (usually studied in the active matter literature) to the passive case (already a very hard and not yet
settled problem). 

Many open questions remain unanswered and pose important challenges. We comment on a few of them below.

The first issue that calls for a careful analysis is how does the phase diagram transform
from the one for disks to the one for dumbbells when one smoothly varies the form of the molecules to interpolate between these two limits.

A careful study of the dynamics of the topological defects and their influence upon the 
phase transitions is definitely needed. Qi {\it et al.} (2014) and Kapfer \& Krauth (2015) suggested that there is a percolation of the 
defect string network at the liquid-hexatic transition of passive models with sufficiently hard potentials, and that  these strings  surround 
domains with hexatic order. Is it the case for the active model as well?

 In equilibrium a duality transformation linking the interacting
defect system relevant for two-dimensional melting to a 
Laplacian roughening model is often used to attack the former by simulating the latter.
However, such a relation does not necessarily hold under active forces since it relies on 
a transformation of the partition function. Are there other transformations of similar kind 
that could be used in the active case?

The hexatic phase is only stable in a minute density regime in the case of hard disks, 
which can be missed very easily in both simulations and experiments.
Additionally, the order of the transition is difficult to ascertain due to finite-size effects.
Therefore, although the picture that we described here is consistent and very attractive, 
it still needs to be confirmed with more detailed numerically simulations and, 
hopefully, experimental measurements. A rigorous proof, as the one developed by 
van Enter and Shlosman (2002) for the planar spin model seems out of reach for the 
active model since it is specific to equilibrium conditions (Gibbs states, partition functions). 
Could any other kind of rigorous proof be worked out for the active case?

\begin{figure}
\begin{center}
\includegraphics[scale=0.2]{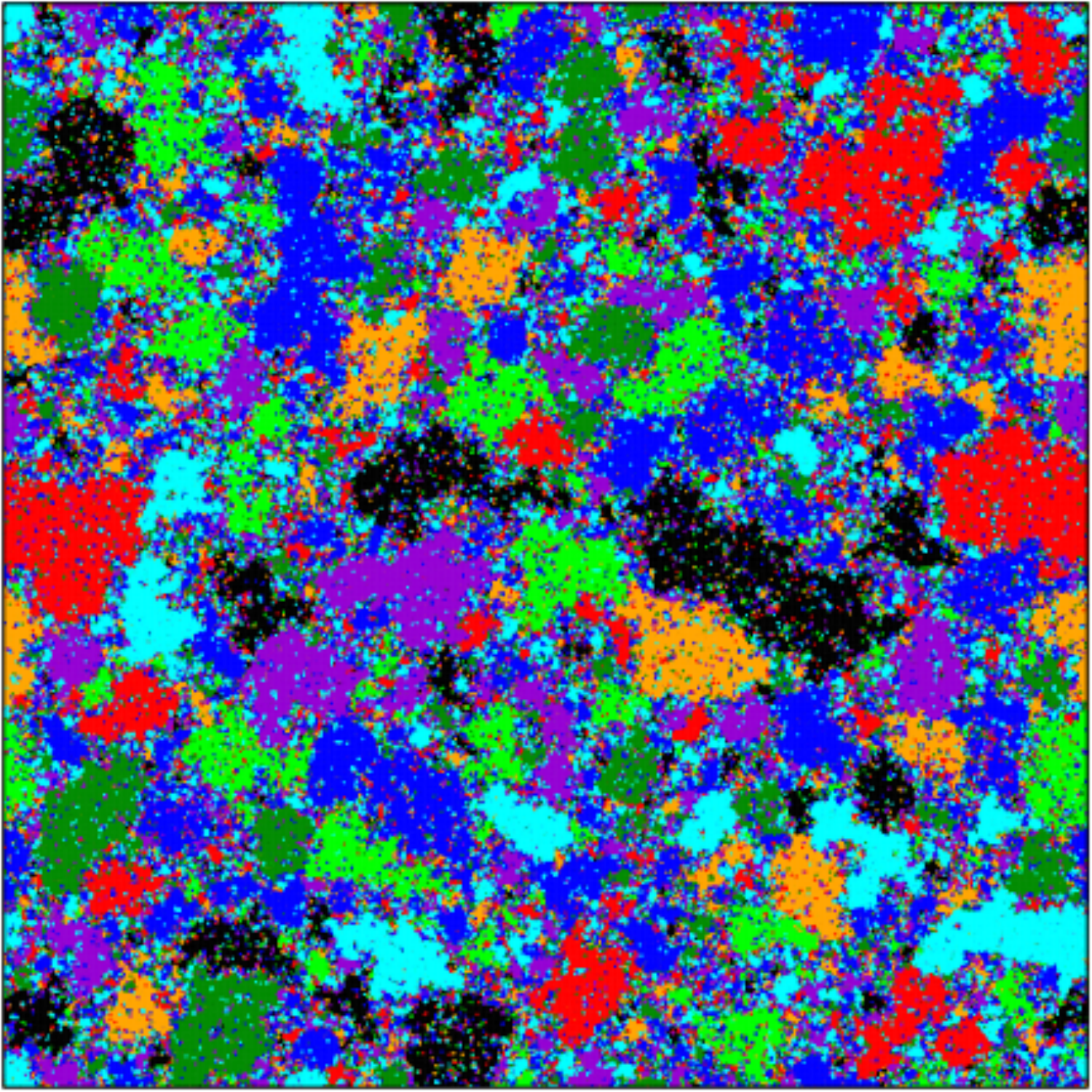}
\includegraphics[scale=0.2]{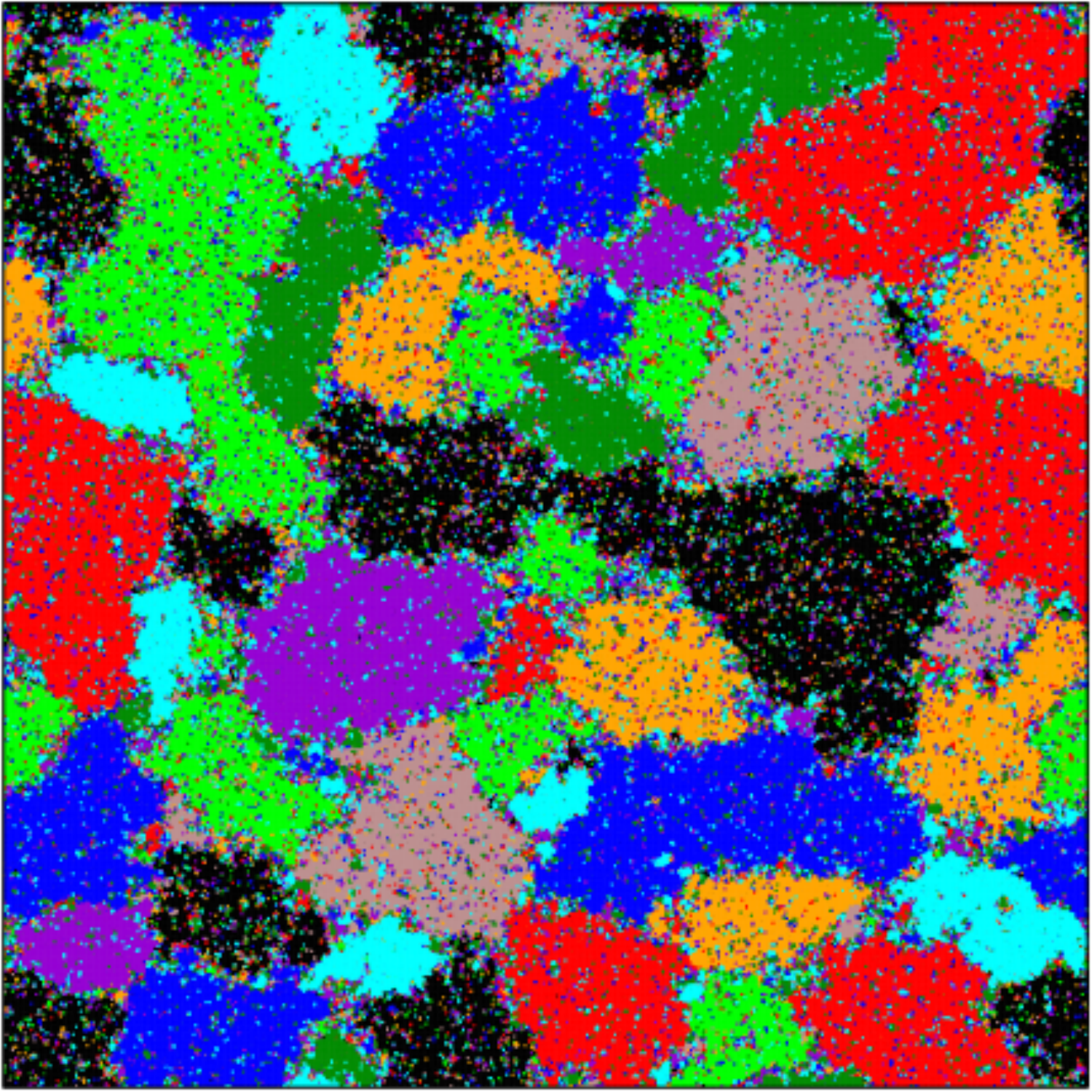}
\includegraphics[scale=0.2]{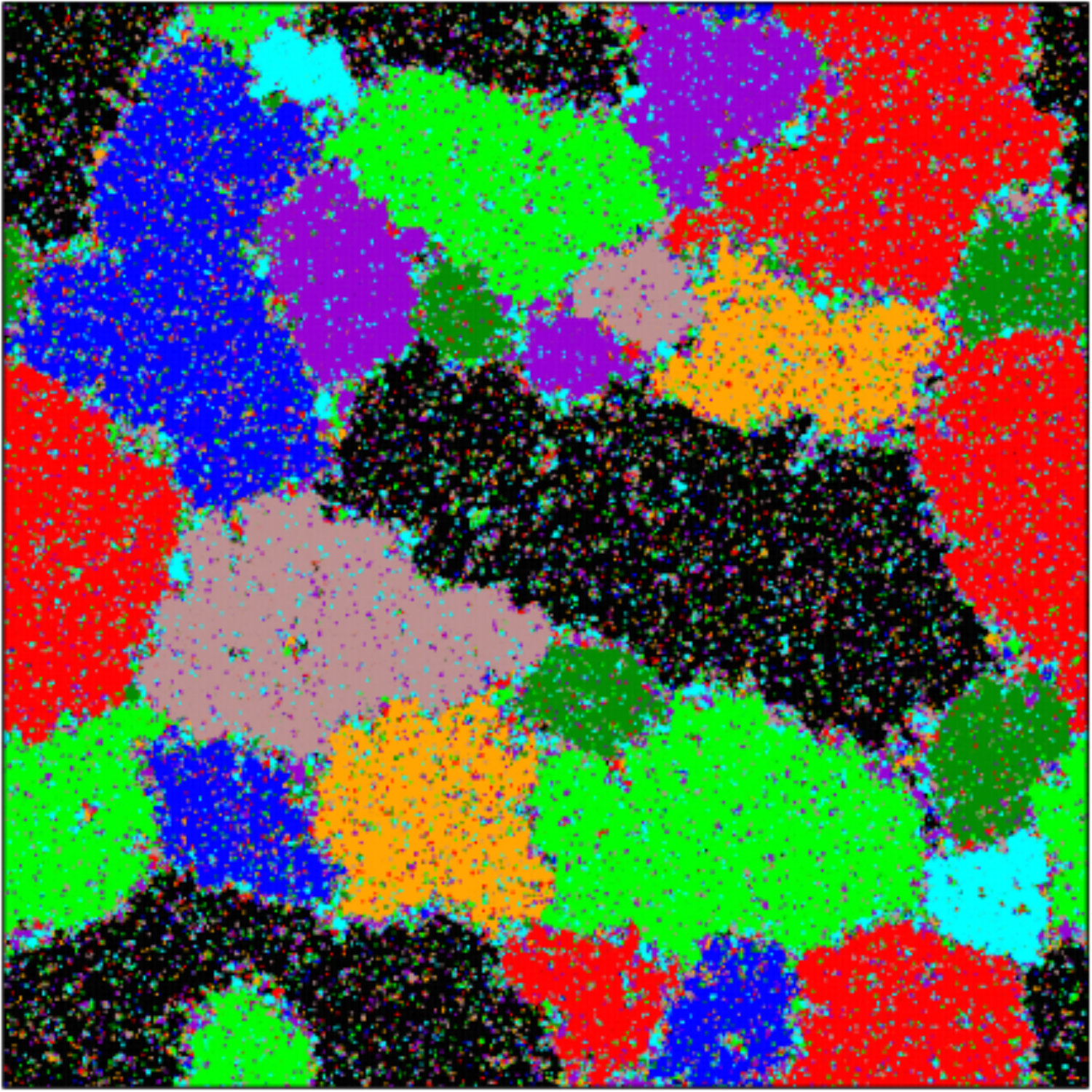}
\includegraphics[scale=0.2]{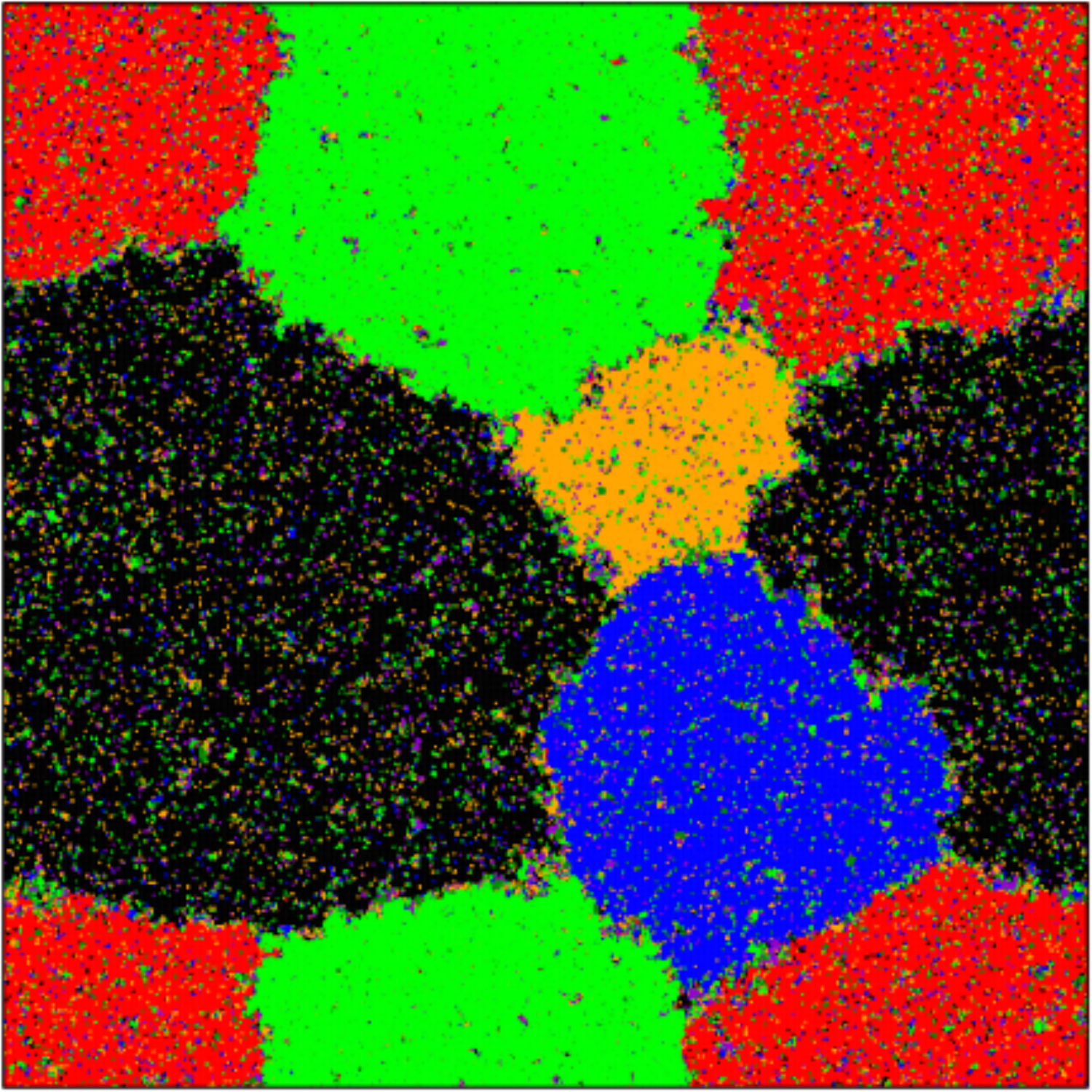}
\end{center}
\caption{Four snapshots taken at subsequent times that are representative of the dynamics following a quench across the first order transition 
of the Potts model with $q=9$. Each colour represents one state among the nine possible ones of the spins. Figure taken from Corberi {\it et al.} (2018).
}
\label{fig:potts9q}
\end{figure}

The dynamics across a first order phase transition occurs {\it via} nucleation of the stable phase into the unstable 
one. This problem is usually discussed with a single state that wins the competition against another one when the 
transition is crossed. A slightly more complex example is the one of the Potts model with $q>4$,
a model in which the stable states in the ordered phase are $q$ degenerate ones, and the relevant 
dynamic process is a multi-nucleation one. If many stable phases of different kind nucleate simultaneously 
these will grow quickly until they touch and block. The further evolution is a normal coarsening one. The time-scales for
nucleation, growth of sufficiently large bubbles and coarsening are very different and can be numerically quantified 
exploiting data from Monte Carlo simulations~\cite{Esposito18}. The dynamics 
across the first order liquid-hexatic transition should have similar features to the ones just described.
This problem is under study in our group. 

\bibliographystyle{OUPnamed}
\bibliography{leshouches-biblio.bib}

\end{document}